\documentclass[12pt]{article}
\usepackage{a4wide}

\usepackage[dvipdfmx,hiresbb]{graphicx} 
\usepackage{mediabb}                    

\usepackage{amsmath,amssymb}
\usepackage{color}
\usepackage[bookmarks,bookmarksnumbered]{hyperref}
\usepackage{cellspace}
\usepackage{tikz}
\usepackage{mathtools}
\usepackage{cite}
\usetikzlibrary{fadings}
\usetikzlibrary{patterns}
\usetikzlibrary{shadows.blur}
\usetikzlibrary{shapes}
\usepackage{ascmac}
\usepackage{enumitem}

\newcommand{\aln}[1]{\begin{align}#1\end{align}}
\newcommand{\nn}{\nonumber\\}

\begin{document}
\title{\vspace{-3cm}
\vbox{
\baselineskip 14pt
\hfill \hbox{\normalsize 
}}  \vskip 1cm
\bf \Large 
Phases and propagation of closed $p$-brane
\vskip 0.5cm
}
\author{
Kiyoharu Kawana\thanks{E-mail: \tt kkiyoharu@kias.re.kr}
\bigskip\\
\normalsize
\it 
School of Physics, Korean Institute for Advanced Study, Seoul 02455, Korea
\smallskip
}
\date{\today}

\maketitle   
\begin{abstract}

We study phases and propagation of closed $p$-brane within the framework of effective field theory with higher-form global symmetries, i.e., {\it brane-field theory}.    
%
%
We extend our previous studies by including the kinetic term of the center-of-mass motion as well as the kinetic term for the relative motions constructed by the area derivatives.    
This inclusion gives rise to another scalar Nambu-Goldstone mode in the broken phase, enriching the phase structures of $p$-brane.   
For example, when the higher-form global symmetries are discrete ones, we show that the low-energy effective theory in the broken phase is described by a topological field theory of the axion $\varphi(X)$ and $p$-form field $A_p^{}(X)$ with multiple (emergent) higher-form global symmetries.  
After the mean-field analysis, we investigate the propagation of $p$-brane in the present framework. 
We find the (functional) plane-wave solutions for the kinetic terms and derive a path-integral representation of the brane propagator. 
This representation motivates us to study the brane propagation within the Born-Oppenheimer approximation, where the volume of $p$-brane is treated as constant.  
In the volume-less limit (i.e. point-particle limit), the propagator reduces to the ordinary propagator of relativistic particle, whereas it describes the propagation of the area elements in the large-volume  limit.  
Correspondingly, it is shown that the Hausdorff dimension of $p$-brane varies from $2$ to $2(p+1)$ as we increase the $p$-brane volume within the Born-Oppenheimer approximation.    
Although these results are quite intriguing, we also point out that the Born-Oppenheimer approximation is invalid in the point-particle limit, highlighting the quantum nature of $p$-brane as an extended object in spacetime.

\end{abstract} 

\setcounter{page}{1} 

\newpage  

\tableofcontents

\section{Introduction}\label{Sec:intro}

%
%

In the recent studies~\cite{Hidaka:2023gwh,Kawana:2024fsn,10.1093/ptep/ptaf023}, we have proposed the effective field theory of closed $p$-brane $C_p^{}$ with higher-form global symmetries~\cite{Gaiotto:2014kfa,Kapustin:2005py,Pantev:2005zs,Nussinov:2009zz,Banks:2010zn,Kapustin:2013uxa,Aharony:2013hda,Kapustin:2014gua,Gaiotto:2017yup,
McGreevy:2022oyu,Brennan:2023mmt,Bhardwaj:2023kri,Luo:2023ive,Gomes:2023ahz,Shao:2023gho}, inspired by the pioneering work~\cite{Iqbal:2021rkn} where string field theory with $1$-form global symmetry was investigated. 
A key point of our construction is making use of the ``{\it area derivative}"~\cite{Eguchi:1979qk,Ogielski:1980ht,Yoneya:1980bw,Banks:1980sq,Migdal:1983qrz,Makeenko:1980vm,Polyakov:1980ca,Kawai:1980qq,PhysRevD.40.3396} for the kinetic term of $p$-brane, which describes a functional variation of the brane field $\phi[C_p^{}]$ under a small $p$-dimensional deformation, $C_p^{}\rightarrow C_p^{}+\delta C_p^{}$.  
Building on the action that is invariant under higher-form global transformations, we have performed the mean-field analysis and demonstrated that various fundamental results  associated with the spontaneous breaking of higher-form symmetries can be systematically  explained as in ordinary Landau field theory for $0$-form symmetries. 
%

In this paper, we extend our study and conduct a more detailed analysis of brane-field theory with a particular focus on the $p$-brane propagation   as well as the mean-field analysis.     
In the first part of Section~\ref{sec:2}, we provide a geometrical definition of the area derivative by considering a discretized space(time) and taking the continuum-limit. 
Readers will see that the area derivative can be interpreted as a natural generalization of the ordinary derivative $\partial_\mu^{}\phi(x)=\lim_{a\rightarrow 0}(\phi(x^\mu+a)-\phi(x^\mu))/a$ with the replacement of $x^\mu$ by a $p$-brane $C_p^{}$ and $a$ with a small $p$-dimensional deformation $\delta C_p^{}$.~(See Eq.~(\ref{def:area derivative}) for the concrete definition.)
Besides, such a geometrical definition of the area derivative indicates  that it is essentially different from the conventional functional derivative $\delta/\delta X^\mu(\xi)$, where the latter can describe  more general functional variations of the brane-field $\phi[C_p^{}]$. 
In this sense, the area derivative can be regarded as a reduced functional derivative of $\delta/\delta X^\mu(\xi)$, but has several good properties compared to it.

After introducing the area derivative, we propose a field-theoretic model of $p$-brane with higher-form global symmetries.  
%
One of the crucial differences compared to the previous studies~\cite{Hidaka:2023gwh,Kawana:2024fsn,10.1093/ptep/ptaf023,Iqbal:2021rkn} is the inclusion of the center-of-mass kinetic term of $p$-brane in the Lagrangian. 
By construction, the area derivative quantifies the functional variations of the brane-field $\phi[C_p^{}]$ induced by a deformation of the relative coordinates of $p$-brane~(see Eq.~(\ref{relative coordinates}) for the definition of the relative coordinates), and it does not contain the information about the center-of-mass motion.  
In the previous studies, this point was overlooked and we solely focused  on the kinetic term constructed by the area derivatives.  
As in ordinary quantum field theory (QFT), the presence of the center-of-mass kinetic term leads to another Nambu-Goldstone (NG) mode $\varphi(x)$ associated with the $0$-form global symmetry in addition to the $p$-form NG mode $A_p^{}(x)$, which then enriches the low-energy structure of the model. 
When the higher-form global symmetries are $\mathrm{U}(1)$, the low-energy effective theory in the broken phase is described by the simple gapless theory of $\varphi(x)$ and $A_p^{}(x)$ while it becomes gapped when these $\mathrm{U}(1)$ global symmetries are explicitly broken down to discrete ones. 
%
The corresponding low-energy effective theory is given by a topological field theory of the axion and $p$-form field with multiple (emergent) higher-form symmetries, resulting in a higher-group structure in general.  
%
%
For $D=4$ and $p=1$, such an effective theory is known as the topological  axion electrodynamics~\cite{Hidaka:2021mml,Hidaka:2024kfx}.

Following the discussion of mean-field analysis, we investigate the propagation of $p$-brane in Section~\ref{sec:3}.   
Compared to a naive model employing the ordinary functional derivative $\delta /\delta X^\mu(\xi)$, the propagation of $p$-brane can be more intuitively analyzed due to the separation of the kinetic terms between  the center-of-mass and relative motions.    
In particular, the (functional) plane-wave solution of the area derivative is obtained by employing the nice differential property~(i.e. Eq.~(\ref{differential property})), which in turn allows us to derive a path-integral expression of the $p$-brane propagator. 
We should note that our path-integral expression resembles the one obtained in the first-quantized approach of $p$-brane~\cite{Ansoldi:1997cw,Ansoldi:2001km,Aurilia:2002aw}. 
%
Although the exact calculation of the path-integral is challenging, our path-integral representation reveals how the complexity of mixed motions appears in the brane action, which then motivates us to study the propagator in the Born-Oppenheimer approximation~\cite{doi:10.1142/9789812795762_0001} by treating the $p$-brane volume $\mathrm{Vol}[C_p^{}]$ as constant.     
%
Within this approximation, we show that the brane propagator reduces to the ordinary propagator of relativistic particle in the volume-less limit (i.e. point-particle limit), whereas it describes the propagation of the area elements $\sigma^{\mu_1^{}\cdots \mu_{p+1}^{}}[C_p^{}]$ in the large-volume  limit.  
As a consistency check, we also argue that the Born-Oppenheimer approximation seems to be valid as long as the proper-time of the brane motion is less than the size of $p$-brane. 
In this sense, the point-particle limit is always invalid, highlighting  the quantum nature of $p$-brane as an extended object in spacetime. 
Furthermore, we estimate the Hausdorff dimension $D_\mathrm{H}^{}$ of $p$-brane within the Born-Oppenheimer approximation and find that it varies from $2$ to $2(p+1)$ as we increase $\mathrm{Vol}[C_p^{}]$. 
While these results are quite intriguing, they also indicate the necessity of exploring the $p$-brane dynamics beyond the Born-Oppenheimer approximation.     
%

\newpage

The organization of this paper is as follows. 
In Sec.~\ref{sec:2}, we provide a geometrical definition of the area derivative and propose a field-theoretic model of closed $p$-brane by utilizing it. 
We then perform the mean-field analysis and show that our brane-field theory can explain a variety of (topological) phases associated with higher-form global symmetries.  
In Sec.~\ref{sec:3}, we explore the propagation of $p$-brane in the present brane-field model and derive the analytic expressions of the propagator within the Born-Oppenheimer approximation.    
Using this analytic results, we also examine the Hausdorff dimension of $p$-brane.  
Section~\ref{sec4} is devoted to summary.  
In Appendix~\ref{app:Functional Derivatives}, we provide additional details on the functional derivative. 
%


\section{Brane field theory}\label{sec:2}

Throughout the paper, we represent a $D$-dimensional spacetime with a metric $g_{\mu\nu}^{}(X)$ by $\Sigma_{D}^{}$ and employ the Minkowski metric signature, $(-,+,+,\cdots,+)$.  
A $p$-dimensional spacelike closed brane is represented by $C_p^{}$, which is given by embedding functions $\{X^\mu(\xi)\}_{\mu=0}^{D-1}$: $S^p\to \Sigma_D^{}$~, where $S^p$ denotes the parameter space of $p$-dimensional sphere whose intrinsic coordinates are $\xi=\{\xi^i\}_{i=1}^p$.   
Besides, we represent the induced metric and its determinant as 
\aln{
h_{ij}^{}(\xi)\coloneq \frac{\partial X^\mu(\xi)}{\partial \xi^i}\frac{\partial X^\nu(\xi)}{\partial \xi^j}g_{\mu\nu}^{}(X(\xi))~,\quad h(\xi)\coloneq {\rm det}(h_{ij}^{})\geq 0~,
} 
which leads to the following expression of the volume of $C_p^{}$:  
\aln{{\rm Vol}[C_p^{}]=\int_{S^p} d^p\xi \sqrt{h(\xi)}=\int_{C_p^{}}E_p^{}~,
\label{volume}
}
where
\aln{
&E_p^{}
=\frac{1}{p!\sqrt{h(\xi)}}
\{X^{\mu_1^{}},\cdots,X^{\mu_p^{}}\}dX_{\mu_1^{}}^{}\wedge \cdots \wedge dX_{\nu_p^{}}^{}~,
\label{volume form}
\\
&\{X^{\mu_1^{}},\cdots,X^{\mu_p^{}}\}=\epsilon^{i_1^{}\cdots i_p^{}}\frac{\partial X^{\mu_1^{}}(\xi)}{\partial \xi^{i_1^{}}}\cdots\frac{\partial X^{\mu_p^{}}(\xi)}{\partial \xi^{i_p^{}}}~.
\label{Nambu bracket}
}
Here, $\epsilon^{i_1^{}\cdots i_p^{}}=\pm 1$ is the totally anti-symmetric tensor and Eq.~(\ref{Nambu bracket}) is the {\it Nambu bracket}~\cite{PhysRevD.7.2405}. 
It satisfies 
\aln{
\{X^{\mu_1^{}},\cdots,X^{\mu_p^{}}\}\{X_{\mu_1^{}}^{},\cdots,X_{\mu_p^{}}^{}\}=p!h(\xi)~,
\label{Ep relation}
}
and 
\aln{\{X^{\mu},X^{\mu_2^{}},\cdots,X^{\mu_p^{}}\}\{X^{\nu},X_{\mu_2^{}}^{},\cdots,X_{\mu_p^{}}^{}\}=(p-1)!h(\xi)\frac{\partial X^\mu(\xi)}{\partial \xi^i}\frac{\partial X^\nu(\xi)}{\partial \xi^j}h^{ij}(\xi)~,
}
where $h^{ij}(\xi)$ is the inverse metric of $h_{ij}^{}(\xi)$. 
%

Our focus in this paper is the complex brane field $\phi[C_p^{}]=\phi[\{X^\mu(\xi)\}]$ which is a functional of $C_p^{}$. 
As in ordinary quantum field theory, we assume that $\phi[C_p^{}]$ is invariant under spacetime diffeomorphism and reparametrization on $C_p^{}$ as 
\aln{
\phi[\{{X'}^{\mu}(\xi)\}]&=\phi[\{X^\mu(\xi)\}]\quad (\text{spacetime diffeomorphism})~,
\\
\phi[\{{X}^{\mu}(\xi')\}]&=\phi[\{X^\mu(\xi)\}]\quad (\text{reparametrization})~.
}
For instance, a functional
\aln{\phi[C_p^{}]=\phi\left(\left\{\int_{C_p^{}}A_p^{(a)}\right\}_a^{}\right)
}
is a typical example that satisfies the above two conditions, where $A_p^{(a)}$ is a general differential $p$-form field. 
%

\subsection{Geometrical functional derivative}\label{Sec:volume integral}

 \begin{figure}
    \centering
    \includegraphics[scale=0.3]{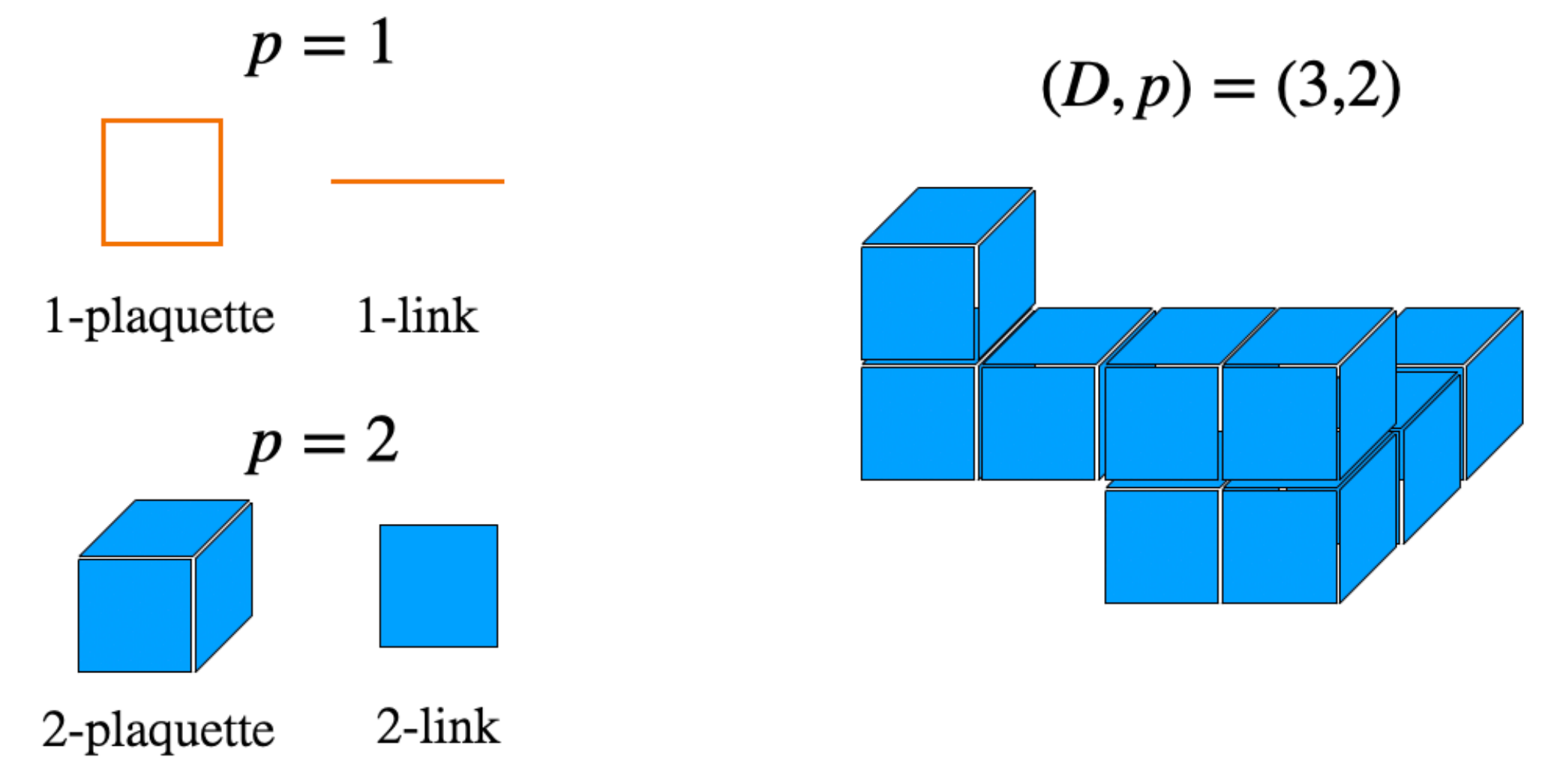}
    \caption{Left: $p$-plaquette and $p$-link in a cubic spacetime lattice. 
    Here we show $p=1$ and $p=2$ cases. 
    \\
    Right: An example of discretized closed $2$-brane. 
    }
    \label{fig:lattice}
\end{figure}
A fundamental issue in constructing a field theory of branes is how to express its kinetic term.  
A general variation of $\phi[C_p^{}]$ under an arbitrary deformation of $p$-brane $\delta C_p^{}=\{\delta X^{\mu}(\xi)\}_{\mu=0}^{D-1}$ is expressed by the conventional functional derivative as
\aln{
\label{general functional derivative}
\delta \phi[C_p^{}]=\int_{S^p} d^p\xi~\delta X^\mu(\xi)\frac{\delta \phi[C_p^{}]}{\delta X^\mu(\xi)}~.
}
See Appendix~\ref{app:Functional Derivatives} for more details about the  functional derivatives. 
Although this expression applies to an arbitrary deformation of a $p$-dimensional manifold, its geometric meaning is less clear except for $p=0$, where the functional derivative reduces to the ordinary derivative $\partial /\partial x^\mu$.   
In the following, we will show that the right-hand side in Eq.~(\ref{general functional derivative}) can be also expressed in terms of geometrical functional derivative known as the {\it area derivative} when $\delta C_p^{}$ is a genuinely $p$-dimensional deformation of $C_p^{}$, i.e. when $C_p^{}+\delta C_p^{}$ preserves the topology of $C_p^{}$. 
A physically intuitive and transparent way to introduce the area derivative  is to consider a discretized spacetime and take the continuum-limit as follows. 
 
 \

Consider a $D$-dimensional cubic lattice $\Lambda_D^{}$ with a lattice spacing $a$. 
In this discretized spacetime, a closed $p$-brane $C_p^{}$ is expressed by  gluing $p$-dimensional minimum hypercubes, as depicted in the right-panel in Fig.~\ref{fig:lattice}. (In this case, a $2$-dimensional surface $C_2^{}$ is obtained by gluing many squares.)
Following the same convention as lattice gauge theory, we call such a minimum $p$-dimensional hypercube {\it $p$-link}. 
Besides, a minimum closed $p$-brane is represented by the boundary of a $(p+1)$-link and we refer to it as {\it $p$-plaquette}. 
They are graphically shown in the left-panel in Fig.\ref{fig:lattice}.
In particular, a $p$-plaquette existing in the $(p+1)$-dimensional subspace $(X^{\mu_1^{}},\cdots,X^{\mu_{p+1}})$ is represented  by $P^{\mu_1^{}\cdots \mu_{p+1}^{}}$. 
%
%
Moreover, we represent the center-of-mass of a given $p$-link as $\hat{i}$ below.   

\begin{figure}
    \centering
    \includegraphics[scale=0.3]{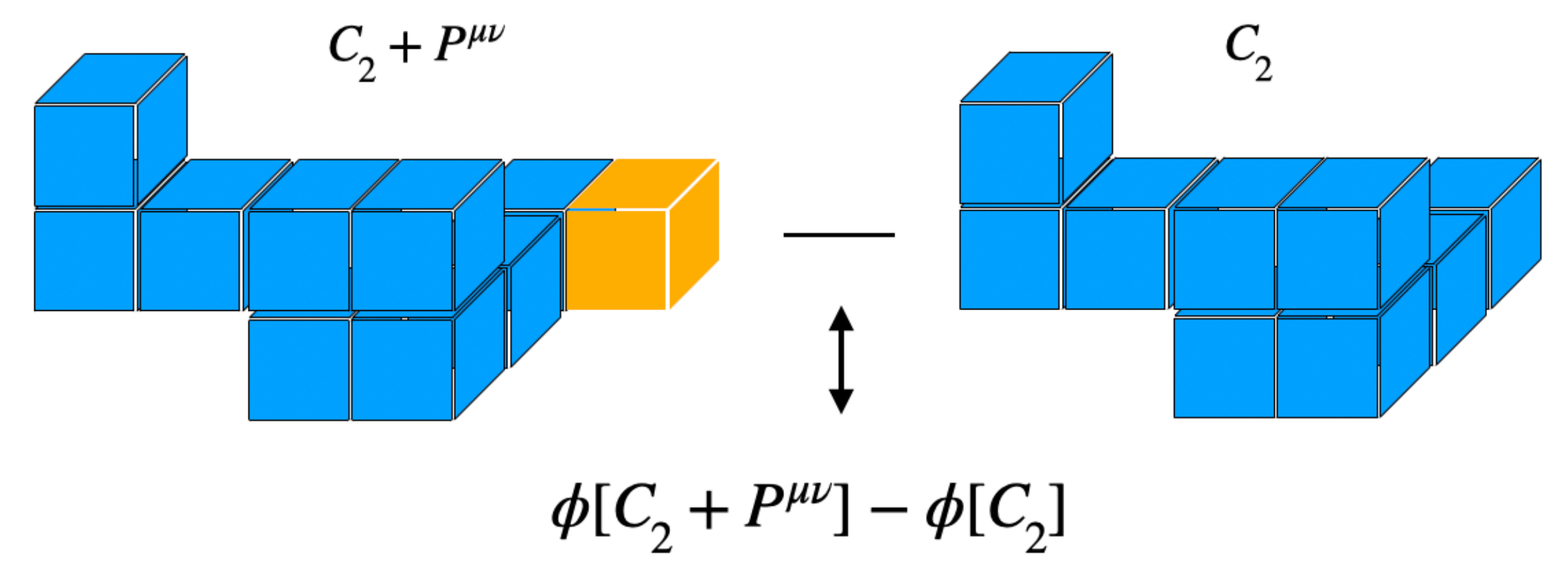}
    \caption{
    A graphical representation of the functional variation of the $2$-brane field $\phi[C_2^{}]$ when adding a $2$-plaquette. 
    }
    \label{fig:area-derivative}
\end{figure}

In this setup, an infinitesimal deformation of $C_p^{}$ is represented by gluing a $p$-plaquette to one of the $p$-links on $C_p^{}$ as illustrated in Fig.~\ref{fig:area-derivative}. 
This addition results in a new $p$-brane $C_p^{}+P^{\mu_1^{}\cdots \mu_{p+1}^{}}(\hat{i})$, where $(\hat{i})$ implies that a $p$-plaquette is glued to the $p$-link whose center-of-mass position is $\hat{i}$. 
By construction, $C_p^{}+P^{\mu_1^{}\cdots \mu_{p+1}^{}}(\hat{i})$ has the same topology as $C_p^{}$.   
Now, we can define geometrical functional derivative via 
\aln{\frac{\delta \phi[C_p^{}]}{\delta \sigma^{\mu_1^{}\cdots \mu_{p+1}^{}}(\hat{i})}\coloneq \frac{\phi[C_p^{}+P^{\mu_1^{}\cdots \mu_{p+1}^{}}(\hat{i})]-\phi[C_p^{}]}{a^{p+1}}~,
\label{def:area derivative}
}  
which is known as the {\it area derivative} for $p=1$~\cite{Migdal:1983qrz,Makeenko:1980vm,Polyakov:1980ca,Kawai:1980qq,PhysRevD.40.3396,Iqbal:2021rkn,Hidaka:2023gwh,Kawana:2024fsn,10.1093/ptep/ptaf023}.   
In the following, we refer to Eq.~(\ref{def:area derivative}) (and its continuum-limit) as the area derivative too for $^\forall p\geq 1$. 
%
%
It should be emphasized that this area derivative is defined only when the topology of $\delta C_p^{}$ is the same as $S^p$ by construction.   
In this sense, the area derivative is a reduced functional derivative compared to $\delta/\delta X^\mu(\xi)$. 

\begin{figure}
    \centering
    \includegraphics[scale=0.4]{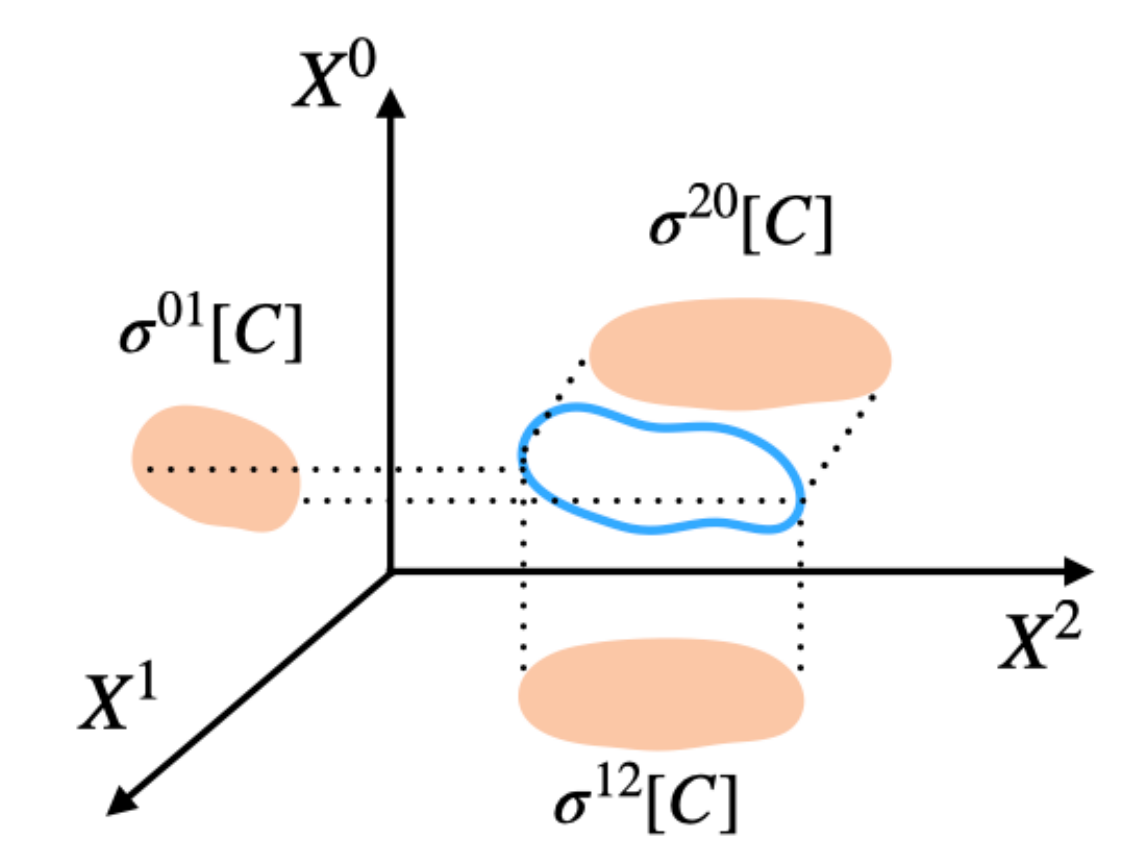}
    \caption{Area elements (orange) of a loop $C$ (blue).  
   }
    \label{fig:area element}
\end{figure}

Equation~(\ref{def:area derivative}) implies that the variation $\delta \phi[C_p^{}]$ can be written as 
\aln{
\delta \phi[C_p^{}]&=\frac{1}{(p+1)!}\int_{\delta D_{p+1}^{}}dX^{\mu_1^{}}\wedge \cdots \wedge dX^{\mu_{p+1}^{}}\frac{\delta \phi[C_p^{}]}{\delta \sigma^{\mu_1^{}\cdots \mu_{p+1}^{}}(\xi)}
\label{variation by area derivative}
\\
&=\frac{1}{(p+1)!}\sigma^{\mu_1^{}\cdots \mu_{p+1}^{}}[\delta C_p^{}(\xi)]\frac{\delta \phi[C_p^{}]}{\delta \sigma^{\mu_1^{}\cdots \mu_{p+1}^{}}(\xi)}
}
in the continuum-limit, where $\delta C_p^{}(\xi)$ is the infinitesimal closed manifold corresponding to $P^{\mu_1^{}\cdots \mu_{p+1}^{}}(\hat{i})$, and 
\aln{
\sigma^{\mu_1^{}\cdots \mu_{p+1}^{}}[C_p^{}]=\int_{D_{p+1}^{}}dX^{\mu_1^{}}\wedge \cdots \wedge dX^{\mu_{p+1}^{}}=\frac{1}{(p+1)!}\int_{C_p^{}} X^{[\mu_1^{}}(\xi)dX^{\mu_2^{}}\wedge \cdots \wedge dX^{\mu_{p+1}^{}]}~
\label{area element}
}
is the area element of the $(p+1)$-dimensional open manifold $D_{p+1}^{}$ surrounded by $C_p^{}$.  
In the last equality, we have used the Stokes theorem and introduced the anti-symmetrization by
\aln{
A^{[\mu_1^{}\mu_2^{}\cdots \mu_p^{}]}\coloneq \sum_{\sigma \in S_{p}^{}}\mathrm{sgn}(\sigma)A^{\sigma(\mu_1^{})\sigma(\mu_2^{})\cdots \sigma(\mu_p^{})}~,
\label{eq:antisymmetrization}
}
where $S_p^{}$ is the symmetric group of degree $p$ and $\mathrm{sgn}(\sigma)$ is the sign of permutation. 
Intuitively, the area element~(\ref{area element}) represents the surface area of the projected $p$-brane onto the $(X^{\mu_1^{}},\cdots ,X^{\mu_{p+1}^{}})$ subspace.  
In Fig.~\ref{fig:area element}, we show an example of $1$-brane.   

%
Equation~(\ref{variation by area derivative}) must coincide with Eq.~(\ref{general functional derivative}), which then provides a relation between $\delta /\delta X^\mu(\xi)$ and $\delta/\delta \sigma^{\mu_1^{}\cdots \mu_{p+1}^{}}(\xi)$ as~\cite{Hidaka:2023gwh}
\aln{
\frac{\delta}{\delta X^\mu(\xi)}=\frac{1}{p!}\{X^{\nu_1^{}},\cdots,X^{\nu_p^{}}\}\frac{\delta}{\delta \sigma^{\mu \nu_1^{}\cdots \nu_p^{}}(\xi)}~ 
\label{relation of functional derivatives}
}
as long as $\delta C_p^{}$ is a genuinely $p$-dimensional deformation of $C_p^{}$. 
For more general deformations of $C_p^{}$, this relation does not necessary  hold and  $\delta /\delta X^\mu(\xi)$ encompasses more functional information. 
In particular, the right-hand side in Eq.~(\ref{relation of functional derivatives}) depends only on the relative coordinates of $p$-brane defined by  
\aln{
Y^\mu(\xi)\coloneq X^\mu(\xi)-x^\mu_{\mathrm{CM}}~,
\label{relative coordinates}
}
where 
\aln{
x_{\mathrm{CM}}^{\mu}\coloneq \frac{1}{\mathrm{Vol}[C_p^{}]}\int_{S^p}d^p\xi \sqrt{h(\xi)}X^{\mu}(\xi)~
\label{CM coordinates}
}
is the center-of-mass coordinate. 
Namely, the right-hand side in Eq.~(\ref{relation of functional derivatives}) cannot describe a variation of the brane-field $\phi[C_p^{}]$ against the change of the center-of-mass position $x_{\mathrm{CM}}^{\mu}~\rightarrow~x_{\mathrm{CM}}^{\mu}+\delta x_{\mathrm{CM}}^{\mu}$. 
Thus, Eq.~(\ref{relation of functional derivatives}) is generalized as 
\aln{
\frac{\delta}{\delta X^\mu(\xi)}=\frac{\partial}{\partial x_\mathrm{CM}^\mu}+\frac{1}{p!}\{X^{\nu_1^{}},\cdots,X^{\nu_p^{}}\}\frac{\delta}{\delta \sigma^{\mu \nu_1^{}\cdots \nu_p^{}}(\xi)}+\cdots 
}
for a general deformation $\delta C_p^{}$. 
Here, $\cdots$ denotes other contributions that cannot be described by the first two terms.   
%

What are the advantages of the area derivative ? 
The most important property is a differential property when acting on a volume integral of a differential $p$-form $A_p^{}(X)$ as~\cite{Hidaka:2023gwh}
\aln{
\frac{\delta}{\delta \sigma^{\mu_1^{}\cdots \mu_{p+1}^{}}(\xi)}\left(\int_{C_p^{}}A_p^{}\right)=F_{\mu_1^{}\cdots \mu_{p+1}^{}}^{}(X(\xi))~,
\label{differential property}
}
where $F_{\mu_1^{}\cdots \mu_{p+1}^{}}^{}(X)$ is the field strength defined by  
\aln{
F_{p+1}^{}(X)=\frac{1}{(p+1)!}F_{\mu_1^{}\cdots \mu_{p+1}^{}}^{}(X)dX^{\mu_1^{}}\wedge \cdots \wedge dX^{\mu_{p+1}^{}}\coloneq dA_p^{}(X)~. 
}
This property plays a crucial role in the following discussion.    
Besides, one can see that the area derivative of the $p$-brane volume is  
\aln{
\frac{\delta \mathrm{Vol}[C_p^{}]}{\delta \sigma^{\mu_1^{}\cdots \mu_{p+1}^{}}(\xi)}=(dE_p^{}(\xi))_{\mu_1^{}\cdots \mu_{p+1}^{}}
}
by Eq.~(\ref{volume form}). 
This implies that the volume derivative is included in the area derivative as 
\aln{
\frac{\delta }{\delta \sigma^{\mu_1^{}\cdots \mu_{p+1}^{}}(\xi)}=(dE_p^{}(\xi))_{\mu_1^{}\cdots \mu_{p+1}^{}}\frac{\partial}{\partial \mathrm{Vol}[C_p^{}]}~.
\label{volume part}
}
See also Refs.~\cite{Migdal:1983qrz,Makeenko:1980vm,Iqbal:2021rkn} for  $p=1$ case.

\subsection{Brane field action}
%
Reflecting the fact that $p$-brane has a lot of dynamical degrees of freedom, we can consider a variety of kinetic terms such as\footnote{$\sqrt{h(\xi)}$ in the denominator comes from the fact that $\delta/\delta X^\mu(\xi)$ is not invariant under the reparametrization on $C_p^{}$.  
See Appendix~\ref{app:Functional Derivatives} for the details.  
}
\aln{
&\int_{S^p}\frac{d^p\xi}{\sqrt{h(\xi)}}~\frac{\delta \phi^*[C_p^{}]}{\delta X^\mu(\xi)}\frac{\delta \phi[C_p^{}]}{\delta X_\mu^{}(\xi)}~,\quad \int_{S^p} d^p\xi\sqrt{-h(\xi)}\frac{\delta \phi^*[C_p^{}]}{\delta \sigma^{\mu_1^{}\cdots \mu_{p+1}^{}}(\xi)}\frac{\delta \phi[C_p^{}]}{\delta \sigma_{\mu_1^{}\cdots \mu_{p+1}^{}}^{}(\xi)}~,\label{various kinetic terms 1}
\\
&\int_{S^p} \frac{d^p\xi}{\sqrt{h(\xi)}}\{X^{\nu_1^{}},\cdots ,X^{\nu_{p}^{}}\}\{X^{\lambda_1^{}},\cdots,X^{\lambda_{p}^{}}\}
\frac{\delta \phi^*[C_p^{}]}{\delta  \sigma^{\mu\nu_1^{}\cdots \nu_{p}^{}}(\xi)}\frac{\delta \phi[C_p^{}]}{\delta {\sigma_\mu^{}}^{\lambda_1^{}\cdots \lambda_{p}^{}}(\xi)}~.
\label{various kinetic terms}
}  
As we mentioned in the previous section, the first one is the most general and encompasses all the functional information of $p$-brane.  
However, such a generality makes the analysis of the model quite challenging as in string field theory. 
In the previous studies~\cite{Hidaka:2023gwh,Kawana:2024fsn}, we have performed the mean-field analysis by assuming the second kinetic term in Eq.~(\ref{various kinetic terms 1}) and found that it can consistently explain the various fundamental results associated with the spontaneous breaking of higher-form global symmetries.      
Note that the third one (\ref{various kinetic terms}) also appears in the string field theory dual to the $\mathrm{SU}(N)$ gauge theory~\cite{10.1093/ptep/ptaf023}. 

In this paper, as a generalization of our previous studies, we  consider the following brane-field action:
\aln{
S[\phi]=-{\cal N}\int [dC_p^{}]\bigg[\frac{1}{\mathrm{Vol}[C_p^{}]^2}\frac{\partial \phi^*[C_p^{}]}{\partial x_{\mathrm{CM}}^\mu}\frac{\partial \phi[C_p^{}]}{\partial x_{\mathrm{CM}\mu}^{}}
&+\frac{1}{\mathrm{Vol}[C_p^{}]}\int_{S^p}d^p\xi \sqrt{-h(\xi)}\frac{\delta \phi^*[C_p^{}]}{\delta \sigma^{\mu_1^{}\cdots \mu_{p+1}^{}}(\xi)}\frac{\delta \phi[C_p^{}]}{\delta \sigma_{\mu_1^{}\cdots \mu_{p+1}^{}}^{}(\xi)}
\nn
&+V(\phi[C_p^{}],\phi^*[C_p^{}])
\bigg]~,
\label{brane-field action}
}
where $V(\phi[C_p^{}],\phi^*[C_p^{}])$ is a general potential, ${\cal N}$ is a normalization factor, and $[dC_p^{}]$ is the (path)integral measure of $C_p^{}$ induced by the diffeomorphism and reparametrization invariant norm
\aln{
||\delta X||^2\coloneq \int_{S^p} d^p\xi \sqrt{-h(\xi)}\delta X^\mu(\xi)\delta X_\mu^{}(\xi)~.
}
In particular, it contains the ordinary spacetime integral of the center-of-mass coordinates 
\aln{
\int_{\Sigma_D^{}}\star 1\in \int [dC_p^{}]~.
} 
Taking this into account, we also express the measure as 
\aln{
\int [dC_p^{}]=\int_{\Sigma_D^{}}\star 1 \int [dY]~,
}
where $[dY]$ denotes the path-integral measure of the relative coordinates. 
In Eq.~(\ref{brane-field action}), the geometrical meaning of the kinetic terms is quite manifest.    
The first term represents the center-of-mass motion of $p$-brane with the inertia factor of $1/\mathrm{Vol}[C_p^{}]^2$, while the second term describes the motion of the area elements $\sigma^{\mu_1^{}\cdots \mu_{p+1}}[C_p^{}]$. 
Although the action~(\ref{brane-field action}) may not capture the full  dynamics of $p$-brane, we will see that it describes the essential aspects of phase structures associated with higher-form global symmetries. 

\subsection{Global symmetries}
Let us discuss global symmetries of the action~(\ref{brane-field action}). 
When the potential $V(\phi[C_p^{}],\phi^*[C_p^{}])$ is a mere function of $\phi^*[C_p^{}]\phi[C_p^{}]$, the action is invariant under the $\mathrm{U}(1)$ $p$-form global transformation
\aln{
\phi[C_p^{}]\quad \rightarrow \quad e^{i\int_{C_p^{}}\Lambda_p^{}}\phi[C_p^{}]~,\quad d\Lambda_p^{}=0~,
\label{p-form transformation}
}
where $\Lambda_p^{}$ is a differential closed (and non-exact) $p$-form. 
In fact, one can check that the kinetic term is invariant under Eq.~(\ref{p-form transformation}) due to Eq.~(\ref{differential property}).  
Besides, the action is invariant under the $\mathrm{U}(1)$ $0$-form global transformation $\phi[C_p^{}]\rightarrow e^{i\theta}\phi[C_p^{}]$, corresponding to the conservation of total number of $p$-branes. 

We can construct the conserved currents by the N\"other method~\cite{Hidaka:2023gwh}. 
Instead of Eq.~(\ref{p-form transformation}), we consider a local $p$-form transformation:
\aln{
\phi[C_p^{}]\quad \rightarrow \quad e^{i\int_{C_p^{}}\Lambda_p^{}}\phi[C_p^{}]~,\quad d\Lambda_p^{}\neq 0~, 
}
which leads to the variation of the action as 
\aln{
\delta S&=-{\cal N}\int [dC_p^{}]\frac{i}{\mathrm{Vol}[C_p^{}](p+1)!}\int_{S^p}d^p\xi\sqrt{h(\xi)}\left(\frac{\delta \phi^*[C_p^{}]}{\delta \sigma^{\mu_1{}^{}\cdots \mu_{p+1}^{}}(\xi)}\phi[C_p^{}]-\phi^*[C_p^{}]\frac{\delta \phi[C_p^{}]}{\delta \sigma^{\mu_{1}^{}\cdots \mu_{p+1}^{}}(\xi)}
\right)
(d\Lambda_p^{})_{\mu_1^{} \cdots \mu_{p+1}^{}}^{}
\nn
&=-\int_{\Sigma_D^{}}d\Lambda_p^{}\wedge \star J_{p+1}^{}~,  
\label{action variation}
}
where 
\aln{
J_{p+1}^{}(X)={\cal N}\int [dC_p^{}]\frac{i}{\mathrm{Vol}[C_p^{}](p+1)!}\int_{S^p} & d^p\xi \sqrt{h(\xi)}\left(\frac{\delta \phi^*[C_p^{}]}{\delta \sigma^{\mu_1{}^{}\cdots \mu_{p+1}^{}}(\xi)}\phi[C_p^{}]-\phi^*[C_p^{}]\frac{\delta \phi[C_p^{}]}{\delta \sigma^{\mu_{1}^{}\cdots \mu_{p+1}^{}}(\xi)}
\right)
\nn
&\times \frac{\delta^{(D)}(X-X(\xi))}{\sqrt{-g(X(\xi))}}
dX^{\mu_1^{}}\wedge \cdots \wedge dX^{\mu_{p+1}^{}}~. 
\label{p+1 form current}
}
When $\phi[C_p^{}]$ satisfies the equation of motion, the variation~(\ref{action variation}) vanishes and we obtain 
\aln{
d\star J_{p+1}^{}(x_\mathrm{CM}^{})=0~,
}
which represents the conservation law of the $\mathrm{U}(1)$ $p$-form global symmetry. 
Similarly, we can consider a local $\mathrm{U}(1)$ $0$-form local  transformation $\phi[C_p^{}]\rightarrow e^{i\theta(x_\mathrm{CM}^{})}\phi[C_p^{}]$ and obtain 
\aln{
J_1^{}(x_\mathrm{CM}^{})={\cal N}&\int [dY]\frac{i}{\mathrm{Vol}[C_p^{}]^2}\left(\frac{\partial \phi^*[C_p^{}]}{\partial x_\mathrm{CM}^{\mu}}\phi[C_p^{}]-\phi^*[C_p^{}]\frac{\partial \phi[C_p^{}]}{\partial x_{\mathrm{CM}}^{\mu}}\right)
dx_\mathrm{CM}^{\mu}~,
\label{0 form current}
} 
which reproduces the ordinary particle current 
\aln{
i(\partial_\mu^{}\phi^*(x_\mathrm{CM}^{})\phi(x_\mathrm{CM}^{})-\phi^*(x_\mathrm{CM}^{})\partial_\mu^{}\phi(x_\mathrm{CM}^{}))dx_\mathrm{CM}^\mu
}
when $p=0$.  

If the potential $V(\phi[C_p^{}],\phi^*[C_p^{}])$ contains terms such as $g\phi[C_p^{}]^N+\mathrm{h.c.}$, these $\mathrm{U}(1)$ global  symmetries are explicitly broken down to $\mathbb{Z}_N^{}$ symmetries given by 
\aln{
\phi[C_p^{}]\quad &\rightarrow \quad e^{\frac{i}{N}\int_{C_p^{}}\Lambda_p^{}}\phi[C_p^{}]\quad \text{with}\quad d\Lambda_p^{}=0~,\quad \int_{C_p^{}}\Lambda_p^{}\in 2\pi \mathbb{Z}~,
\\
\phi[C_p^{}]\quad &\rightarrow \quad e^{\frac{2\pi i}{N}n}\phi[C_p^{}]~,\quad n\in \mathbb{Z}~.
}
In this case, the system is gapped after the spontaneous symmetry breaking and exhibits topological order as we will see below. 
%
%

\subsection{Mean-field analysis}
Here, we perform the mean-field analysis of the theory~(\ref{brane-field action}). 
Since many parts of the calculations overlap with our previous studies~\cite{Hidaka:2023gwh,Kawana:2024fsn,Iqbal:2021rkn}, we will focus on highlighting the main results and differences compared to these previous studies.   
%

As in ordinary QFT, the center-of-mass kinetic term does not play a  crucial role when investigating the vacuum state, which then leads to the following equation of motion of the brane-field:
\aln{
\frac{1}{(p+1)!}\int_{S^p}d^p\xi \frac{\delta}{\delta \sigma^{\mu_1^{}\cdots \mu_{p+1}^{}}(\xi)}\frac{\sqrt{h(\xi)}}{\mathrm{Vol}[C_p^{}]}\frac{\delta \phi[C_p^{}]}{\delta \sigma_{\mu_1^{}\cdots \mu_{p+1}^{}}^{}(\xi)}-\frac{\delta V(\phi[C_p^{}],\phi^*[C_p^{}])}{\delta \phi^*[C_p^{}]}=0~.
\label{equation of motion}
}

\

\

\noindent{\bf UNBROKEN PHASE}\\
Let us consider the unbroken phase of the global symmetries, i.e. we focus on the parameter space such that the potential $V(\phi[C_p^{}],\phi^*[C_p^{})$ has an absolute minimum at $\phi[C_p^{}]=0$. 
In this case, Eq.~(\ref{equation of motion}) admits a solution that exhibits the area law in the large-volume limit, $\mathrm{Vol}[C_p^{}]\rightarrow \infty$~\cite{Hidaka:2023gwh,Kawana:2024fsn,10.1093/ptep/ptaf023,Iqbal:2021rkn}.
In fact, we consider the following ansatz
\aln{\phi[C_p^{}]=\frac{1}{\sqrt{2}}f(\mathrm{Vol}[M_{p+1}^{}])~,
}  
where $f(z)$ is an arbitrary real function and $M_{p+1}^{}$ is the $(p+1)$-dimensional minimal surface enclosed by $C_{p}^{}$, i.e. $\partial M_{p+1}^{}=C_p^{}$. 
Analogously to Eq.~(\ref{volume}), we represent the minimum-surface volume as 
\aln{
\mathrm{Vol}[M_{p+1}^{}]=\int_{M_{p+1}^{}}E_{p}^{}=\frac{1}{(p+1)!}\int_{M_{p+1}^{}}(E_{p+1}^{})_{\mu_1^{}\cdots \mu_{p+1}^{}}dX^{\mu_1^{}}\wedge \cdots \wedge dX^{\mu_{p+1}^{}}~.
}
Putting the above ansatz into Eq.~(\ref{equation of motion}), we obtain the effective equation of motion:
\aln{
f''(z)-q(z)f'(z)-\frac{\delta V(f)}{\delta f(z)}=0~,\quad z=\mathrm{Vol}[M_{p+1}^{}]~,
\label{eom under ansatz}
}
where 
\aln{
q(z)=\frac{1}{(p+1)!}\int_{S^p}d^p\xi\frac{\delta}{\delta \sigma^{\mu_1^{}\cdots \mu_{p+1}^{}}(\xi)}\left(\frac{\sqrt{h(\xi)}}{\mathrm{Vol}[C_p^{}]}E^{\mu_1^{}\cdots \mu_{p+1}^{}}(\xi)\right)~,
}
which is dependent only on the geometrical information of $C_p^{}$ and does not depend on the parameters of $V(\phi[C_p^{}],\phi^*[C_p^{}])$. 
Thus, it should behave as $q(z)\propto z^{-1}$ for $z\rightarrow \infty$ on dimensional ground and Eq.~(\ref{eom under ansatz}) can be approximated as  
\aln{f''(z)-T_p^2f(z)\approx 0
\label{eom in large z}
}
for $z\rightarrow \infty$, where we have expanded the potential around $\phi[C_p^{}]=0$ as $V(\phi[C_p^{}],\phi^*[C_p^{}])\approx T_p^2 \phi[C_p^{}]\phi^*[C_p^{}]$. 
Equation~(\ref{eom in large z}) corresponds to the area law solution 
\aln{
\phi[C_p^{}]\approx \frac{1}{\sqrt{2}}\exp\left(-T_p^{}\times \mathrm{Vol}[M_{p+1}^{}]\right)\quad \text{for } z\rightarrow \infty~,
}
which is a characteristic behavior of the unbroken phase of higher-form symmetries. 
One can see that $T_p^{}$ corresponds to the $p$-brane tension.

\

\noindent{\bf BROKEN PHASE}\\
The global symmetries are spontaneously broken when the brane-field develops a nonzero VEV, $\langle \phi[C_p^{}]\rangle=v/\sqrt{2}\neq 0$. 
The Nambu-Goldstone modes are given by the phase modulations defined by 
\aln{
\phi[C_p^{}]=\frac{v}{\sqrt{2}}\exp\left(i\varphi(x_{\mathrm{CM}}^{})+i\int_{C_p^{}}A_p^{}\right)~,
\label{phase modulations}
}
where $\varphi(x_{\mathrm{CM}}^{})$ is a compact real scalar field satisfying $\varphi(x_{\mathrm{CM}}^{})\sim \varphi(x_{\mathrm{CM}}^{})+2\pi$, and $A_p^{}$ is a $p$-form field. 
From the point of view of these phase modulations, the original global transformations correspond to 
\aln{\varphi \quad &\rightarrow \quad \varphi+\theta~,\quad \theta \in \mathbb{R}~,
\label{varphi transformation}
\\
A_{p}^{}\quad &\rightarrow \quad A_p^{}+\Lambda_p^{}~,\quad d\Lambda_p^{} =0~,
\label{Ap transformation}
} 
and the low-energy effective theory must be invariant under these transformations. 
In fact, we can explicitly calculate it as~\cite{Hidaka:2023gwh,Kawana:2024fsn}
\aln{
S_{\mathrm{eff}}^{}[\varphi,A_p^{}]=-\frac{v^2}{2}\int_{\Sigma_D^{}} \left[d\varphi\wedge \star d\varphi+F_{p+1}^{}\wedge \star F_{p+1}^{}
\right]~,
\label{effective theory}
}
where $F_{p+1}^{}=dA_p^{}$ is the field strength of $A_p^{}$ and $\star$ is the Hodge dual operator. 
The first term is the ordinary kinetic term of the NG mode $\varphi$, whereas the second term corresponds to the $p$-form Maxwell theory.   

 Equation~(\ref{effective theory}) has emergent (magnetic) higher-form symmetries represented by  
 \aln{
 dd\varphi=0~,\quad dF_{p+1}^{}=0~,
 }
 which correspond to $\mathrm{U}(1)$ $(D-2)$- and $(D-p-2)$-form global symmetries respectively.  
The corresponding charges are 
\aln{
Q_{D-2}^{}&=\frac{1}{2\pi}\int_{C_1^{}}d\varphi \quad \in  \mathbb{Z}~,\quad Q_{D-p-2}^{}=\frac{1}{2\pi}\int_{C_{p+1}^{}}F_{p+1}^{}\quad \in  \mathbb{Z}~.
}
These emergent higher-form symmetries indicate the existence of $(D-2)$- and $(D-p-2)$-dimensional topological defects in spacetime. 
%
The former is nothing but the well-known global vortex configuration and the latter corresponds to a topologically non-trivial $(D-p-3)$-dimensional static configuration characterized by the winding number of $C_p^{}\sim S^p$ on $S^{p+1}$ for a given time slice~\cite{Kawana:2024qmz}. 
%

\

\noindent {\bf DISCRETE SYMMETRIES AND TOPOLOGICAL ORDER}\\
When the $\mathrm{U}(1)$ global symmetries are explicitly broken down to $\mathbb{Z}_N^{}$ by the explicit breaking terms such as $\phi[C_p^{}]^N$, the system is no longer gapless in the broken phase and exhibits topological order. 
By repeating the same calculations as  Refs.~\cite{Hidaka:2023gwh,Kawana:2024fsn}, the low-energy effective action in the presence of the $\phi[C_p^{}]^N$ potential term is given as 
\aln{
S_{\mathrm{eff}}^{}=\frac{N}{2\pi }\int_{\Sigma_{D}^{}}H_{D-1}^{}\wedge d\varphi &+\frac{N}{2\pi }\int_{\Sigma_{D}^{}} B_{D-p-1}^{}\wedge dA_p^{}~,
\label{BF theory}
}
where $H_{D-1}^{}(X)$ and $B_{D-p-1}^{}(X)$ are $(D-1)$- and $(D-p-1)$-form fields. 
This is a $\mathrm{BF}$-type topological field theory and exhibits topological order.   
In addition to the original $\mathbb{Z}_N^{}$ $0$- and $p$-form symmetries~(\ref{varphi transformation})(\ref{Ap transformation}), Eq.~(\ref{BF theory}) has emergent $\mathbb{Z}_N^{}$ $(D-1)$- and $(D-p-1)$-form symmetries expressed by 
\aln{
H_{D-1}^{}\quad &\rightarrow \quad H_{D-1}^{}+\frac{1}{N}\Lambda_{D-1}^{}~,\quad d\Lambda_{D-1}=0~,\quad \int_{C_{D-1}^{}}\Lambda_{D-1}^{}\in 2\pi \mathbb{Z}~,
\\
B_{D-p-1}^{}\quad &\rightarrow \quad B_{D-p-1}^{}+\frac{1}{N}\Lambda_{D-p-1}^{}~,\quad d\Lambda_{D-p-1}=0~,\quad \int_{C_{D-p-1}^{}}\Lambda_{D-p-1}^{}\in 2\pi \mathbb{Z}~,
}
where $C_{D-1}^{}$~($C_{D-p-1}^{}$) is a $(D-1)~((D-p-1))$-dimensional closed subspace.
Namely, Eq.~(\ref{BF theory}) has totally four topological operators:
\aln{
U_0^{}(C_{D-1}^{})&=\exp\left(i\int_{C_{D-1}^{}}H_{D-1}^{}\right)~,\quad U_p^{}(C_{D-p-1}^{})=\exp\left(i\int_{C_{D-p-1}}B_{D-p-1}^{}\right)~,
\\
U_{D-1}^{}(P,P')&=\exp\left(i(\varphi(P)-\varphi(P'))\right)~,\quad U_{D-p-1}^{}(C_p^{})=\exp\left(i\int_{C_p^{}}A_p^{}\right)~, 
} 
where $P$ and $P'$ denote spacetime points.  
When the spatial manifold is flat, i.e. $\Sigma_{D-1}^{}=\mathbb{R}^{D-1}$, all the topological operators (except for $U_{D-1}^{}(P,P')$) can trivially shrink to a point, which then implies $\langle U_0^{}(C_{D-1}^{})\rangle=\langle U_p^{}(C_{D-p-1}^{})\rangle=\langle U_{D-p-1}^{}(C_p^{})\rangle=1$. 
Since they also correspond to the charged objects under these higher-form symmetries, this means that these symmetries are spontaneously broken.  
%

\begin{figure}
    \centering
    \includegraphics[scale=0.35]{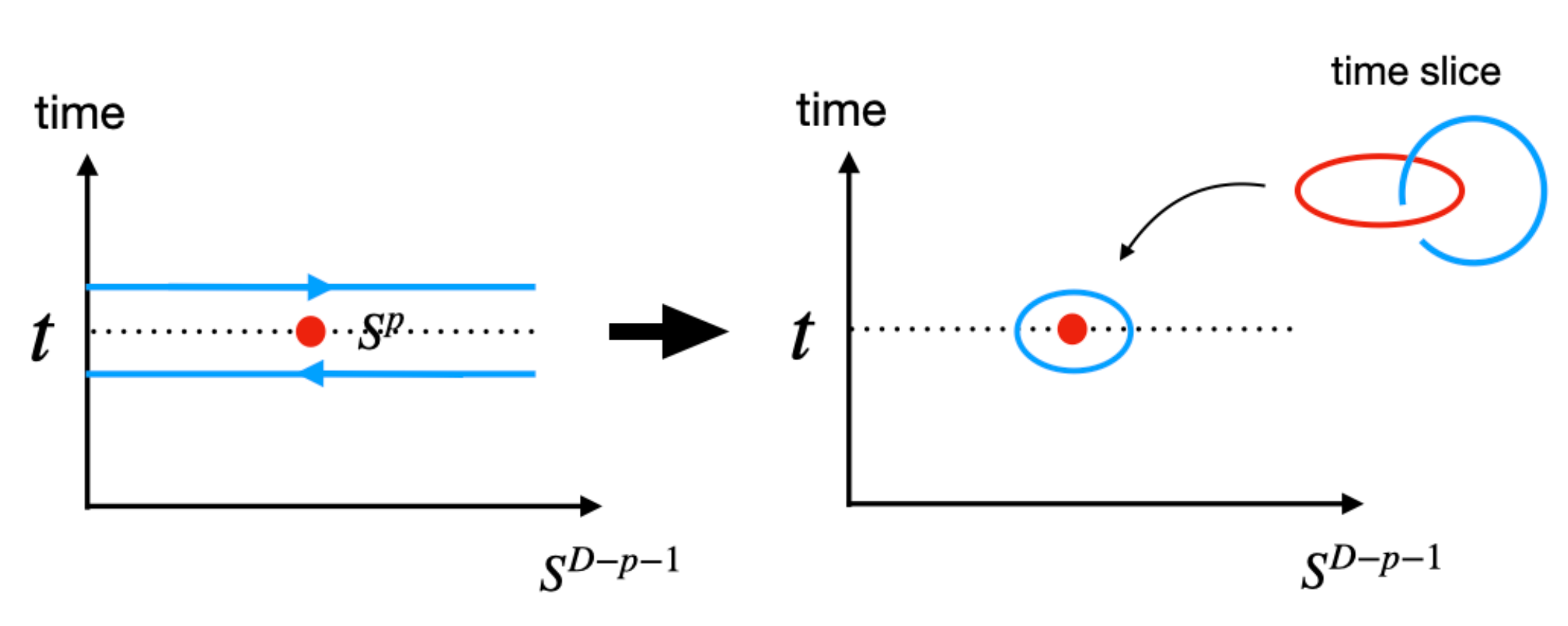}
    \caption{Graphical representation of the operator relation~(\ref{commutation relation}). 
   }
    \label{fig:commutation}
\end{figure}
On the other hand, when the topology of spatial manifold is nontrivial, these topological operators can act nontrivially on the ground states. 
As an example, let us consider $\Sigma_{D-1}^{}=S^{p}\times S^{D-p-1}$ and choose $C_{p}=S^p$ and $C_{D-p-1}^{}=S^{D-p-1}$.   
In this case, two symmetry operators $U_p^{}(S^{D-p-1})$ and $U_{D-p-1}^{}(S^p)$ are linked and satisfies 
\aln{
U_p^{}(S^{D-p-1})U_{D-p-1}^{}(S^p)U_{p}^{-1}(S^{D-p-1})=e^{i\frac{2\pi }{N}}U_{D-p-1}^{}(S^p)
\label{commutation relation}
}
as an equal-time operator relation.
In Fig.~\ref{fig:commutation}, we show the graphical representation of this relation. 
%
Then, when we choose a ground state $|\Omega \rangle$ as an simultaneous  eigenstate of $U_p^{}(S^{D-p-1})$ as 
\aln{
U_p^{}(S^{D-p-1})|\Omega\rangle=e^{i\theta}|\Omega \rangle~,\quad \theta \in \mathbb{R}~,
}
$|\Omega'\rangle\coloneq U_{D-p-1}^{}(S^p)|\Omega \rangle$ also has the same energy as $|\Omega\rangle$ because $U_{D-p-1}^{}(S^p)$ is a symmetry operator. 
However, it has a different eigenvalue of $U_p^{}(S^{D-p-1})$ as 
\aln{U_p^{}(S^{D-p-1})|\Omega'\rangle=e^{i\theta+i\frac{2\pi}{N}}|\Omega'\rangle
}
due to the commutation relation~(\ref{commutation relation}), which implies that $|\Omega '\rangle$ is another degenerate ground state. 
By repeating the same calculations, one can see that there are $N$-degenerate states in this case. 
%


\subsection{More on phases of higher-form symmetries}

So far, we have considered brane-field theories with simple contact interactions.    
The present effective framework, however, also allows a variety of interactions that can lead to diverse (quantum) phases of branes in the low-energy limit.   
For example, we can consider non-local interactions such as~\cite{Yoneya:1980bw,Iqbal:2021rkn} 
\aln{\int [dC_p^{(1)}]\int [dC_p^{(2)}]\int [dC_p^{(3)}]\delta(C_p^{(1)}+C_p^{(2)}+C_p^{(3)})\phi[C_p^{(1)}]\phi[C_p^{(2)}]\phi[C_p^{(3)}]+\mathrm{h.c.}~,
}
where $\delta(C_p^{(1)}+C_p^{(2)}+C_p^{(3)})$ is a delta function in the $p$-brane space that restricts the path-integral configurations to $C_p^{(1)}=-C_p^{(2)}-C_p^{(3)}$. 
This interaction preserves the $\mathrm{U}(1)$ $p$-form global symmetry as 
\aln{
&\delta(C_p^{(1)}+C_p^{(2)}+C_p^{(3)})\phi[C_p^{(1)}]\phi[C_p^{(2)}]\phi[C_p^{(3)}]
\nn
\rightarrow \quad &\delta(C_p^{(1)}+C_p^{(2)}+C_p^{(3)})e^{i\int_{C_p^{(1)}+C_p^{(2)}+C_p^{(3)}}\Lambda_p^{}}
\phi[C_p^{(1)}]\phi[C_p^{(2)}]\phi[C_p^{(3)}]
\nn
&=\delta(C_p^{(1)}+C_p^{(2)}+C_p^{(3)})\phi[C_p^{(1)}]\phi[C_p^{(2)}]\phi[C_p^{(3)}]~,
}
while it explicitly breaks the $\mathrm{U}(1)$ $0$-form global symmetry down to $\mathbb{Z}_3^{}$. 
In this case, the $p$-form phase modulation $A_p^{}$ remains gapless after the symmetry breaking, whereas the axion field $\varphi(x_\mathrm{CM}^{})$ is gapped as we have seen in the previous sections. 
Furthermore, we can consider the higher area-derivative terms such as 
%
%
\aln{ 
i\lambda\int [dC_p^{}] 
\frac{1}{\mathrm{Vol}[C_p^{}]}\int_{S^p}d^p\xi \sqrt{h(\xi)}~D_{p+1}^{}\phi[C_p^{}]\wedge \cdots \wedge D_{p+1}^{}\phi[C_p^{}]
+\mathrm{h.c.}~,
\label{another type of explicit breaking}
}
for $D=(p+1)M~,~M\in \mathbb{Z}$, where $\lambda\in \mathbb{R}$ is a coupling constant and 
\aln{
D_{p+1}\phi[C_p^{}]\coloneq \frac{1}{(p+1)!}\frac{\delta \phi[C_p^{}]}{\delta \sigma^{\mu_1^{}\cdots \mu_{p+1}^{}}(\xi)}dX^{\mu_1^{}}\wedge \cdots \wedge dX^{\mu_{p+1}^{}}~.
}
Equation~(\ref{another type of explicit breaking}) explicitly breaks the $\mathrm{U}(1)$ global symmetries down to $\mathbb{Z}_M^{}$ and  
%
%
%
results in the following low-energy effective term in the broken phase:
\aln{
\lambda\int_{\Sigma_D^{}}
\sin\left(M\varphi\right)
F_{p+1}^{}\wedge \cdots \wedge F_{p+1}^{}~,
} 
where we have neglected the higher-dimensional terms of $A_p^{}$ for simplicity.
%
%
Thus, when the potential $V(\phi[C_p^{}],\phi^*[C_p^{}])$ explicitly breaks the $\mathrm{U}(1)$ global symmetries down to $\mathbb{Z}_N^{}$, the full low-energy effective action becomes
\aln{
S_{\mathrm{eff}}^{}\approx \frac{N}{2\pi}\int_{\Sigma_{D}^{}}H_{D-1}^{}\wedge d\varphi +\frac{N}{2\pi}\int_{\Sigma_{D}^{}} B_{D-p-1}^{}\wedge dA_p^{}
+\lambda\int_{\Sigma_D^{}}
\sin\left(M\varphi\right)
F_{p+1}^{}\wedge \cdots \wedge F_{p+1}^{}~.
\label{TAE}
}
This is a similar effective action to the topological axion electrodynamics~\cite{Hidaka:2021mml,Hidaka:2024kfx}.\footnote{
In the topological axion electrodynamics, the interaction between $\varphi$ and $A_1^{}$ is given by the anomaly interaction $\frac{N}{8\pi^2}\int_{\Sigma_4^{}}\varphi F_2^{}\wedge F_2^{}$ where the integer $N$ is determined by an axion model. 
In the present model, however, $\lambda$ can take any (real) values. 
} 
In the present model, the remaining global symmetries are four higher-form symmetries described by the following equations of motion:  
\begin{itemize}
\item $\mathbb{Z}_q^{}$ $0$-form symmetry: $\frac{N(-1)^D}{2\pi }dH_{D-1}^{}+\lambda M \cos (M\varphi)F_{p+1}^{}\wedge \cdots \wedge F_{p+1}^{}=0$
\item $\mathbb{Z}_q^{}$ $p$-form symmetry: $\frac{N (-1)^{D-p-1}}{2\pi}dB_{D-p-1}^{}+
\lambda d\sin(M\varphi)\wedge F_{p+1}^{}\wedge \cdots \wedge F_{p+1}^{}=0$
\item $\mathbb{Z}_N^{}$ $(D-p-1)$-form symmetry: $dA_p^{}=0$
\item $\mathbb{Z}_N^{}$ $(D-1)$-form symmetry: $d\varphi=0$~,
\end{itemize}
where $q$ is the greatest common divisor between $N$ and $M$.
Each equations of motion can be interpreted as a topological conservation law $d\star J_{n+1}^{}=0$, giving a topological charge $Q_n^{}\coloneq \int_{C_{D-n-1}^{}}\star J_{n+1}^{}$ where $C_{D-n-1}^{}$ is a $(D-n-1)$-dimensional subspace.   
As discussed in Ref.~\cite{Hidaka:2021mml,Hidaka:2024kfx}, this theory exhibits the topological order on the axionic domain wall by the intersections of $0$- and $p$-form symmetry generators, in addition to the bulk topological order discussed in the previous section,
%
%
Besides, these four higher-form symmetries give rise to the structure of higher-group symmetry where the correlations (or intersections) between symmetry operators satisfy nontrivial relations as quantum operators.

\    

\noindent{\bf GAUGING HIGHER-FORM GLOBAL SYMMETRIES}\\
It is also possible to gauge the higher-form global symmetries~\cite{Hidaka:2023gwh,Kawana:2024fsn}. 
The gauged version of Eq.~(\ref{brane-field action}) in a differential form  is 
\aln{
S_\text{gauged}^{}=&-{\cal N}\int [dC_p^{}]\bigg[\frac{1}{\mathrm{Vol}[C_p^{}]^2}(D_{\mathrm{CM}}^{}\phi[C_p^{}])^*\wedge \star D_\mathrm{CM}^{}\phi[C_p^{}]
\nn
&+\frac{1}{\mathrm{Vol}[C_p^{}]}\int_{S^p}d^p\xi \sqrt{-h(\xi)}(D_\sigma^{}\phi[C_p^{}])^*\wedge \star D_\sigma^{}\phi[C_p^{}]
+V(\phi[C_p^{}]\phi^*[C_p^{}])
\bigg]
\nn
&-\frac{1}{2g^2}\int_{\Sigma_D^{}}F_{2}^{}\wedge \star F_{2}^{}-\frac{1}{2\tilde{g}^2}\int_{\Sigma_D^{}}F_{p+2}^{}\wedge \star F_{p+2}^{}~,
\label{gauged action}
}
where $g^2$ and $\tilde{g}^2$ are the gauge couplings and 
\aln{
D_\mathrm{CM}^{}\phi[C_p^{}]&=(d-iq_0^{}A_1^{}(x_{\mathrm{CM}}^{}))\phi[C_p^{}]~,
\\
D_\sigma^{}\phi[C_p^{}]&=\frac{1}{(p+1)!}\left(\frac{\delta}{\delta \sigma^{\mu_1^{}\cdots \mu_{p+1}}(\xi)}-iq_p^{}A_{\mu_1^{}\cdots \mu_{p+1}^{}}(X(\xi))\right)\phi[C_p^{}]dX^{\mu_1^{}}\wedge \cdots \wedge dX^{\mu_{p+1}^{}}~,
\\
&F_2^{}=dA_1^{}~,\quad F_{p+2}^{}=dA_{p+1}^{}~
}
with $q_0^{},q_{p}^{}\in \mathbb{Z}$. 
The gauge transformations are given by 
\aln{\phi[C_p^{}]\quad \rightarrow \quad e^{iq_0^{}\Lambda(x_\mathrm{CM}^{})}\phi[C_p^{}]~,\quad A_1^{}(x_\mathrm{CM}^{})\quad \rightarrow \quad  A_1^{}(x_\mathrm{CM}^{})+d\Lambda(x_\mathrm{CM}^{})~,
\\
\phi[C_p^{}]\quad \rightarrow \quad e^{iq_p^{}\int_{C_p^{}}\Lambda_p^{}}\phi[C_p^{}]~,\quad A_{p+1}^{}(X)\quad \rightarrow \quad  A_{p+1}^{}(X)+d\Lambda_{p}(X)~,
}
where $\Lambda(x_\mathrm{CM}^{})$ ($\Lambda_p^{}(X)$) is a differential $0~(p)$-form. 
The equations of motion of the gauge fields are
\aln{
\frac{1}{g^2}d\star F_2^{}&=q_0^{}\star J_1^{}~,\quad \frac{(-1)^p}{\tilde{g}^2}d\star F_{p+2}^{}=q_p^{}\star J_{p+1}^{}~,
}
by which we obtain the on-shell conservation laws, $d\star J_1^{}=0~,~d\star J_{p+1}^{}=0$. 
The corresponding conserved charges are 
\aln{
Q_0^{}&=\int_{\Sigma_{D-1}}\star J_1^{}=\frac{1}{g^2q_0^{}}\int_{C_{D-2}^{}}\star F_2^{}\in \mathbb{Z}~,
\\
 Q_p^{}&=\int_{\Sigma_{D-p-1}}\star J_{p+1}^{}=\frac{1}{\tilde{g}^2q_p^{}}\int_{C_{D-p-2}^{}}\star F_{p+2}^{}\in \mathbb{Z}~,
}
where $\Sigma_{D-1}^{}$ ($\Sigma_{D-p-1}^{}$) is an $(D-1)$ ($(D-p-1)$)-dimensional open subspace with a closed boundary $\partial \Sigma_{D-1}^{}=C_{D-2}^{}$ ($\partial \Sigma_{D-p-1}^{}=C_{D-p-2}^{}$). 
These conserved charges mean the existence of the electric $\mathbb{Z}_{q_0^{}}^{}$ $1$-form and $\mathbb{Z}_{q_p^{}}^{}$ $(p+1)$-form global symmetries, whose transformations are given by
\aln{
A_1^{}&\quad \rightarrow \quad A_1^{}+\frac{1}{q_0^{}}\Lambda_1^{}~,\quad d\Lambda_1^{}=0~,\quad \int_{C_1^{}}\Lambda_1^{}\in 2\pi \mathbb{Z}~,
\label{1-form sym}
\\
A_{p+1}^{}&\quad \rightarrow \quad A_{p+1}^{}+\frac{1}{q_p^{}}\Lambda_{p+1}^{}~,\quad d\Lambda_{p+1}^{}=0~,\quad \int_{C_{p+1}^{}}\Lambda_{p+1}^{}\in 2\pi \mathbb{Z}~,
\label{p+1-form sym}
}
where $C_1^{}$~($C_{p+1}^{}$) is a $1$~($(p+1)$)-dimensional closed subspace. 
In addition, the theory has the magnetic $\mathrm{U}(1)$ $(D-3)$- and $(D-p-3)$-form global symmetries described by
\aln{
dF_2^{}=0~,\quad dF_{p+2}^{}=0~.
\label{magnetic symmetries}
}
In the coulomb phase, the brane-field exhibits the area law $\langle \phi[C_p^{}]\rangle\sim \exp(-T_p^{}\mathrm{Vol}[M_{p+1}^{}])$ as shown in the previous section, and the low-energy effective theory is simply described by the Maxwell theories of two gauge fields $A_1^{}$ and $A_{p+1}^{}$.   
These gapless modes can be interpreted as the Nambu-Goldstone modes of the broken magnetic higher-form symmetries~(\ref{magnetic symmetries})~\cite{Kawana:2024fsn}.  
%

In the superconducting (Higgs) phase, i.e. $\langle \phi[C_p^{}]\rangle=v/\sqrt{2}\neq 0$, on the other hand, the gauge fields become massive and the low-energy effective theory is described by a $\mathrm{BF}$-type topological field theory~\cite{Hidaka:2023gwh,Kawana:2024fsn}.   
Here, we present the essence of the derivation based on the symmetry argument. 
See Refs.~\cite{Hidaka:2023gwh,Kawana:2024fsn} for the detailed derivation.   
By introducing the phase fluctuations as Eq.~(\ref{phase modulations}), one can check that the classical currents~(\ref{p+1 form current})(\ref{0 form current}) become 
\aln{
J_1^{}(x)\propto v^2d\varphi(x)~,\quad 
J_{p+1}^{}(x)\propto v^2dA_p^{}(x)~,
}
which satisfies $dJ_1^{}=dJ_{p+1}^{}=0$, representing the emergent $\mathbb{Z}_{q_0^{}}^{}$ $(D-2)$- and $\mathbb{Z}_{q_p}^{}$ $(D-p-2)$-form global symmetries in the superconducting phase. 
The corresponding charged objects are the $(D-2)$ $((D-p-2))$-dimensional topological defect which can be analytically constructed in the present theory~\cite{Kawana:2024fsn}.   
The low-energy effective theory must respect these global symmetries as well as the original symmetries~(\ref{1-form sym})(\ref{p+1-form sym}) and is gapped.  
Then, one can easily guess that the effective theory with these requirements is   
\aln{
S_{\mathrm{eff}}^{}=\frac{q_0^{}}{2\pi}\int_{\Sigma_D^{}}B_{D-2}^{}\wedge dA_1^{}+\frac{q_p^{}}{2\pi}\int_{\Sigma_D^{}}B_{D-p-2}^{}\wedge dA_{p+1}^{}~,
}
where $B_{D-2}^{}(X)$ ($B_{D-p-2}^{}(X)$) is the dual field of $\varphi(X)$ ($A_p^{}(X)$). 
This is a $\mathrm{BF}$-type topological field theory and exhibits topological order as explained in the previous section.  

\

In this way, our brane-field theory provides an effective framework for describing a variety of (topological) phases of matter and can be regarded as a natural generalization of conventional Landau field theory for $0$-form symmetries to higher-form symmetries.

\section{Propagation of $p$-brane}\label{sec:3}
In this section, we study the propagation of $p$-brane in the brane-field theory. 
We will derive a path-integral expression of the brane propagator in the Schwinger's proper time formalism and obtain approximate analytic expressions of the brane propagator in the Born-Oppenheimer approximation.  
We also discuss the Hausdorff dimension of $p$-brane within the approximation.

\subsection{Plane wave and propagator}

In the following, we consider the free theory, i.e. 
\aln{
V(\phi[C_p^{}],\phi^*[C_p^{}])=T_p^2\phi^*[C_p^{}]\phi[C_p^{}]~,
\label{free potential}
} 
where $T_p^{}$ corresponds to the $p$-brane tension with a mass dimension $p+1$.  
The time-ordered propagator is formally defined by 
\aln{
G[C_p^{},C_p']=\langle T\{\phi[C_p^{}]\phi^\dagger[C_p^{'}]\}\rangle\coloneqq\frac{1}{Z}\int {\cal D}\phi^* {\cal D}\phi~\phi[C_p^{}]\phi^*[C_p^{'}]e^{iS[\phi]}~,
}
which can be written in the operator form as 
\aln{
=-\frac{i}{2}\langle C_p^{}|\left(\hat{H}-i\varepsilon/2\right)^{-1}|C_p^{'}\rangle=\frac{1}{2}\int_0^{\infty}dA\langle C_p^{}|e^{-i(\hat{H}-i\frac{\varepsilon}{2})A}|C_p^{'}\rangle~,
\label{eq:brane propagator}
}
where $i\varepsilon$ is the Feynman's prescription and 
\aln{
\hat{H}&=\frac{1}{2}\left[-\frac{1}{\mathrm{Vol}[C_p^{}]^2}\frac{\partial^2}{\partial x_{\mathrm{CM}}^\mu \partial x_{\mathrm{CM}\mu}}-\int_{S^p} d^p\xi~\frac{\delta}{\delta \sigma^{\mu_1^{}\cdots \mu_{p+1}^{}}(\xi)}\frac{\sqrt{h(\xi)}}{\mathrm{Vol}[C_p^{}]}\frac{\delta}{\delta \sigma_{\mu_1^{}\cdots \mu_{p+1}^{}}^{}(\xi)}+T_p^2\right]
\label{world-manifold Hamiltonian}
}
is the Hamiltonian on the world-manifold. 
Note that $AT_p^{}$ is interpreted as the proper surface-area of the $(p+1)$-dimensional world-manifold.   
%
%

In order to calculate the above amplitude, we have to find the plane-wave solutions.   
As for the center-of-mass motion, it is well-known as 
\aln{\langle x_{\mathrm{CM}}^{}|k_{\mathrm{CM}}^{}\rangle=\frac{1}{(2\pi)^{D/2}}e^{ik_{\mathrm{CM}}^{}\cdot x_{\mathrm{CM}}^{}}~. 
}
We can similarly find the plane-wave for the area derivative as   
\aln{
\exp\left(i\int_{C_p^{}}K_p^{}\right)\coloneqq \exp\left(\frac{i}{(p+1)!}\int_{C_{p}^{}}K_{\mu_1^{}\cdots \mu_{p+1}^{}}^{}(\xi)Y^{[\mu_1^{}}(\xi)dY^{\mu_2^{}}\wedge \cdots \wedge dY^{\mu_{p+1}^{}]}
\right)~,
\label{plane wave solution}
} 
where $K_{\mu_1^{}\cdots \mu_{p+1}^{}}^{}(\xi)$ is a real anti-symmetric function and we have explicitly used the relative coordinates $\{Y^\mu(\xi)\}$ in order to show that it is independent of $\{x_\mathrm{CM}^\mu \}$.  
Acting the area derivative, we can explicitly check that 
\aln{
-i\frac{\delta}{\delta \sigma^{\mu_1^{}\cdots \mu_{p+1}^{}}(\xi)}\exp\left(i\int_{C_p^{}}K_p^{}\right)
&=(dK_p^{})_{\mu_1^{}\cdots \mu_{p+1}^{}}(\xi)\exp\left(i\int_{C_p^{}}K_p^{}\right)
\nn
&=K_{\mu_1^{}\cdots \mu_{p+1}^{}}^{}(\xi)\exp\left(i\int_{C_p^{}}K_p^{}\right)
}
by Eq.~(\ref{differential property}).    
In particular, the plane-wave of the volume dynamics can be identified by  
\aln{
 (K_p^{\mathrm{Vol}})_{\mu_1^{}\cdots \mu_{p+1}^{}}^{}\coloneq -(dE_p^{}(\xi))_{\mu_1^{}\cdots \mu_{p+1}^{}}^{}\times k_{\mathrm{Vol}}^{}~,\quad k_{\mathrm{Vol}}^{}\in \mathbb{R}~
}
due to Eq.~(\ref{volume part}). 
In fact, by putting this into Eq.~(\ref{plane wave solution}), one can  check 
\aln{
e^{i\int_{C_p^{}}K_p^{\mathrm{Vol}}}=e^{ik_{\mathrm{Vol}}\mathrm{Vol}[C_p^{}]}~
}
with the use of the Stokes theorem. 
In the following, we redefine $K_p^{}-K_p^{\mathrm{Vol}}$ as $K_p^{}$, which satisfies  
\aln{K_p^{}\cdot dE_p^{}\coloneq \frac{1}{(p+1)!}K_{\mu_1^{}\cdots \mu_{p+1}}^{}(dE_p^{})^{\mu_1^{}\cdots \mu_{p+1}}=0~. 
\label{constraint on K}
}
Namely, $K_p^{}$ corresponds to the eigenmode of the shape deformation of $p$-brane.   
Then, the completeness relation is given by 
\aln{
1=\int\frac{d^Dk_\mathrm{CM}^{}}{(2\pi)^D}\int\frac{dk_\mathrm{Vol}^{}}{2\pi}\int [dK_p^{}]\bigg(|k_\mathrm{CM}^{}\rangle \otimes  |k_\mathrm{Vol}^{}\rangle\otimes  |K_p^{}\rangle\bigg)\bigg(
\langle K_p^{}|\otimes \langle k_\mathrm{Vol}^{}| \otimes \langle k_\mathrm{CM}^{}|\bigg)~,
\label{complete relation}
}
with an appropriate normalization, where the (path)integral measure $[dK_p^{}]$ is chosen in the same manner as $[dC_p^{}]$.   
Note that, although $|k_\mathrm{CM}^{}\rangle\otimes |k_\mathrm{Vol}^{}\rangle \otimes |K_p^{}\rangle$ is the eigenstate of the center-of-mass and area derivatives, it is not an eigenstate of the Hamiltonian $\hat{H}$ due to the explicit $\{X^\mu(\xi)\}$ dependence via $h(\xi)$ and $\mathrm{Vol}[C_p^{}]$ in Eq.~(\ref{world-manifold Hamiltonian}).   
By repeatedly inserting the completeness relations into Eq.~(\ref{eq:brane propagator}), we obtain the path-integral expression of the transition amplitude as 
\aln{
\langle C_p^{}|e^{-i(\hat{H}-i\varepsilon)A}|C_p^{'}\rangle=\int_{t=0,C'_p}^{t=A,C_p^{}} {\cal D}X\int {\cal D}k_{\mathrm{CM}}^{}\int {\cal D}k_{\mathrm{Vol}}^{}
\int {\cal D}K^{} e^{iS_p^{}}~,
\label{path-integral amplitude}
}
where 
\aln{
S_p^{}&=\int_0^Adt\bigg(k_{\mathrm{CM}\mu}^{}(t)\dot{x}_{\mathrm{CM}}^\mu (t)+k_{\mathrm{Vol}}^{}(t)\dot{\mathrm{Vol}}[C_p^{}(t)]
-(H(t)-i\varepsilon)
\bigg)+\int_{M_{p+1}^{}}K_{p+1}^{}
\label{world-manifold action}
}
is the first quantized $p$-brane action. 
The following are the definitions of all the quantities appearing in Eq.~(\ref{world-manifold action}):
\aln{
H(t)&=\frac{1}{2}\left(\frac{k^2}{\mathrm{Vol}[C_p^{}(t)]^2}+D[C_p^{}(t)]k_\mathrm{Vol}^{}(t)^2+K_p^{}(t)^{2}+T_p^2\right)~,
\label{energy}
\\
D[C_p^{}(t)]&=\frac{1}{\mathrm{Vol}[C_p^{}(t)]}\int_{S^p}d^p\xi\sqrt{h(t,\xi)} (dE_p^{})\cdot (dE_p^{})
\label{def D}
\\
K_p^{}(t)^2&=\frac{1}{\mathrm{Vol}[C_p^{}(t)]}\int_{S^p}d^p\xi\sqrt{h(t,\xi)} K_p^{}\cdot K_p^{}~,
\label{area energy}
\\
\int_{M_{p+1}}K_{p+1}^{}&=\frac{1}{(p+1)!}\int_{M_{p+1}}K_{\mu_1^{}\cdots \mu_{p+1}^{}}(t,\xi)dY^{\mu_1^{}}\wedge \cdots \wedge dY^{\mu_{p+1}^{}}~,
\label{momentum term}
}
where the product $A\cdot B$ is similarly defined by Eq.~(\ref{constraint on K}) and $M_{p+1}^{}$ is the $(p+1)$-dimensional world-manifold of $p$-brane with boundaries $\partial M_{p+1}^{}=C_p^{}-C_{p}^{'}$. 

%
The path-integral of $\{x_\mathrm{CM}^{\mu}(t)\}$ can be performed as they appear only at the first term in Eq.~(\ref{world-manifold action}), which then produces the boundary terms and the constraint   
\aln{
\dot{k}_{\mathrm{CM}\mu}^{}(t)=0\quad \therefore~ k_{\mathrm{CM}\mu}^{}(t)=k_\mu^{}=\text{constant}~. 
\label{center-of-mass momentums}
} 
This is of course a consequence of spacetime-translation invariance.  
%
%
%
Then, Eq.~(\ref{path-integral amplitude}) becomes  
\aln{
\langle C_p^{}|e^{-i(\hat{H}-i\varepsilon)A}|C_p^{'}\rangle=\int\frac{d^Dk}{(2\pi)^D}e^{ik\cdot (x_{\mathrm{CM}}^{}-x_{\mathrm{CM}}^{'})}
\int_{t=0,\{Y'(\xi)\}}^{t=A,\{Y(\xi)\}} {\cal D}Y\int {\cal D}k_\mathrm{Vol}^{}\int {\cal D}K~e^{i\int_{M_{p+1}}K_{p+1}-i\int_0^A dt(H(t)-i\varepsilon)}~. 
\label{transition amplitude}
}
where we have also decomposed the boundary $p$-branes as 
\aln{
C_p^{}=(\{x_{\mathrm{CM}}^\mu\},\{Y^\mu(\xi)\})~,\quad C_{p}^{'}=(\{{x'}_{\mathrm{CM}}^\mu\},\{{Y'}^\mu(\xi)\})~. 
\label{decomposition of boundary X}
}
Except for $p=0$, the remaining path-integral cannot be evaluated analytically without any further approximations or assumptions.   
In the next section, we consider the Born-Oppenheimer approximation in order to examine Eq.~(\ref{transition amplitude}) analytically. 
%

\begin{figure}
    \centering
    \includegraphics[scale=0.4]{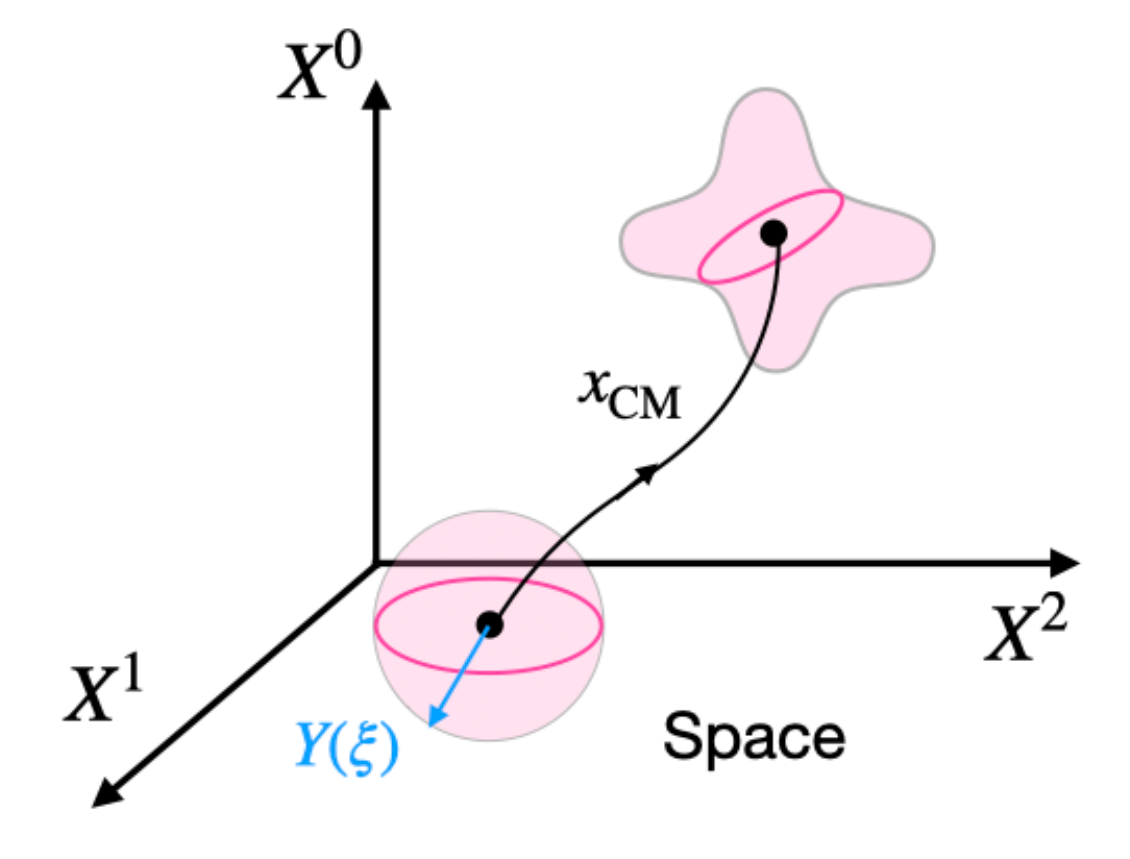}
    \caption{
  An illustration of the propagating $p$-brane in spacetime.  
   }
    \label{fig:propagation}
\end{figure}

\subsection{Born-Oppenheimer treatment}

In general, the motion of $p$-brane is a complex mixture of the center-of-mass and relative motions as illustrated in Fig.~\ref{fig:propagation}. 
This situation reminds us molecular dynamics consisting of atomic nuclei and electrons.  
In particular, when we closely look at the equations (\ref{energy})-(\ref{momentum term}), one can recognize that one of the complexities  of the brane motion originates in the existence of $\mathrm{Vol}[C_p^{}(t)]$ in each terms in the Lagrangian.   
In other words, the complexity seems to be significantly reduced when we  separate the volume-dynamics from the others. 
Such a treatment resembles the Born-Oppenheimer approximation in molecule systems~\cite{doi:10.1142/9789812795762_0001} and we consider a similar treatment in the $p$-brane dynamics below.   
%

We neglect $\dot{\mathrm{Vol}}[C_p^{}(t)]$ in Eq.~(\ref{world-manifold action}) and the second term in Eq.~(\ref{energy}), and represent a fixed $p$-brane volume as  
\aln{
\mathrm{Vol}[C_p^{}]=V_p^{}\quad \leftrightarrow \quad M_{p}^{}[C_p^{}]\coloneq T_p^{}\mathrm{Vol}[C_p^{}]=T_p^{}V_p^{}=M_p^{}~,
}
where $M_p^{}$ can be interpreted as the ``mass" of $p$-brane. 
The validity of this Born-Oppenheimer treatment is discussed in the last part of this section. 
Within this approximation, one can see that the center-of-mass transition amplitude is decoupled from the relative motions in Eq.~(\ref{transition amplitude}) and described by 
\aln{
\langle C_p^{}|e^{-i(\hat{H}-i\varepsilon)A}|C_p^{'}\rangle\bigg|_{\mathrm{CM}}^{}=
\int\frac{d^Dk_\mathrm{CM}^{}}{(2\pi)^D}e^{ik_\mathrm{CM}^{}\cdot (x_{\mathrm{CM}}^{}-x_{\mathrm{CM}}^{'})-i\frac{AT_p^2}{2}\left(\frac{k_\mathrm{CM}^2}{M_p^2}-i\varepsilon\right)}~,
\label{particle transition}
}
 which coincides with the transition amplitude of relativistic particle with a proper time $AT_p^{}/V_p^{}$ and mass $M_p^{}$. 

As for the relative motions, we can at least find a self-consistent saddle-path in the following way.  
First, Eq.~(\ref{momentum term}) can be deformed as 
\aln{
&\int_{M_{p+1}}K_{p+1}^{}=\frac{1}{(p+1)!}\int_{M_{p+1}}\bigg[d(K_{\mu_1^{}\cdots \mu_{p+1}^{}}(t,\xi)Y^{\mu_1^{}})\wedge \cdots \wedge dY^{\mu_{p+1}^{}}
\nn
&\hspace{6cm}-(dK_{\mu_1^{}\cdots \mu_{p+1}^{}}(t,\xi))Y^{[\mu_1^{}}dY^{\mu_2^{}}\wedge \cdots \wedge dY^{\mu_{p+1}^{}]}\bigg]
\nn
=&\int_{C_p^{}}K_p^{}(A)-\int_{C_p^{'}}K_p^{}(0)-\frac{1}{(p+1)!}\int_{M_{p+1}}(dK_{\mu_1^{}\cdots \mu_{p+1}^{}}(t,\xi))Y^{[\mu_1^{}}dY^{\mu_2^{}}\wedge \cdots \wedge dY^{\mu_{p+1}^{}]}
\label{deformed momentum term}
}
by the Stokes theorem. 
Then, if we neglect the $\{Y^\mu(t,\xi)\}$ dependence in Eq.~(\ref{area energy}) (i.e. $h(t,\xi)$), one can see that their dependence appears  only in the last term in Eq.~(\ref{deformed momentum term}) in the exponent in the transition amplitude~(\ref{transition amplitude}), leading to the following saddle-path condition 
\aln{dK_{\mu_1^{}\cdots \mu_{p+1}^{}}(t,\xi)=0\quad \therefore K_{\mu_1^{}\cdots \mu_{p+1}^{}}^{}(t,\xi)=k_{\mu_1^{}\cdots \mu_{p+1}^{}}^{}=\text{constant}~,
\label{global modes}
}  
as well as the center-of-mass momenta~(\ref{center-of-mass momentums}). 
By putting this into Eq.~(\ref{area energy}), one can check 
\aln{
K_p^{}(t)^2=k\cdot k\coloneq \frac{1}{(p+1)!}k_{\mu_1^{}\cdots \mu_{p+1}^{}}^{}k^{\mu_1^{}\cdots \mu_{p+1}^{}}~,
}
which is indeed independent of $\{Y^\mu(\xi)\}$. 
Thus, Eq.~(\ref{global modes}) can be regarded as one of the saddle-paths of  the relative motions, although the existence of other saddle-paths   cannot be ruled out. 

Now the transition amplitude~(\ref{transition amplitude}) in the Born-Oppenheimer and saddle-path approximations can be written as   
\aln{
&\langle C_p^{}|e^{-i\hat{H}A}|C_p^{'}\rangle|_{\mathrm{BO}}^{}
\nn
\approx &\int\frac{d^{D}k_\mathrm{CM}^{}}{(2\pi)^{D}}\int\frac{d^{D'}k}{(2\pi)^{D'}}\exp\left(ik_\mathrm{CM}^{}\cdot (x_\mathrm{CM}^{}-x_{\mathrm{CM}}^{'})+i k\cdot (\sigma[C_p^{}]-\sigma[C_p^{'}])-i\frac{AT_p^2}{2}\left(\frac{k_\mathrm{CM}^2}{M_p^{2}}+\frac{k\cdot k}{T_p^2}+1-i\varepsilon \right)
\right),
\label{approximate amplitude}
}
where $D'={}_D C_{p+1}^{}$ and
\aln{
k\cdot \sigma[C_p^{}]\coloneq \frac{1}{(p+1)!}k_{\mu_1^{}\cdots \mu_{p+1}^{}}\sigma^{\mu_1^{}\cdots \mu_{p+1}^{}}[C_p^{}]~.
}    
Compared to the transition amplitude of relativistic particle~(\ref{particle transition}), Eq.~(\ref{approximate amplitude}) describes the propagation of the area elements $\sigma^{\mu_1^{}\cdots \mu_{p+1}^{}}[C_p^{}]$ of $p$-brane as well.  
Note that the result~(\ref{approximate amplitude}) does not have any dependences on bulk quantities (fluctuations) in the present approximations.   

Although a true propagator of $p$-brane would be more complicated than Eq.~(\ref{approximate amplitude}), it tells us how the motion of $p$-brane with a fixed-volume can vary according to the change of $T_p^{}$ and $V_p^{}$.  
For example, when 
\aln{T_p^{}\rightarrow \infty \quad \text{with}\quad M_p^{}~=~\text{fixed}\quad (\text{Point-particle limit})~,
\label{point-particle limit}
}
Eq.~(\ref{approximate amplitude}) becomes
\aln{
\langle C_p^{}|e^{-i\hat{H}A}|C_p^{'}\rangle|_{\mathrm{BO}}^{}
\approx &\delta^{(D')}(\sigma[C_p^{}]-\sigma[C_p^{'}])\int\frac{d^{D}k_\mathrm{CM}^{}}{(2\pi)^{D}}e^{ik_\mathrm{CM}^{}\cdot (x_\mathrm{CM}^{}-x_{\mathrm{CM}}^{'})-i\frac{AT_p^2}{2}\left(\frac{k_\mathrm{CM}^2}{M_p^{2}}+1-i\varepsilon \right)},
}
which represents the propagation of relativistic particle.  
On the other hand, for
\aln{V_p^{}\rightarrow \infty \quad \text{with} \quad T_p^{}~=~\text{fixed}\quad (\text{Large-brane limit})~,
\label{large brane limit}
}
Eq.~(\ref{approximate amplitude}) becomes
\aln{
\langle C_p^{}|e^{-i\hat{H}A}|C_p^{'}\rangle|_{\mathrm{BO}}^{}
\approx &\delta^{(D)}(x_\mathrm{CM}^{}-x_\mathrm{CM}^{'})\int\frac{d^{D'}k}{(2\pi)^{D'}}e^{i k\cdot (\sigma[C_p^{}]-\sigma[C_p^{'}])-i\frac{A}{2}\left(k\cdot k+T_p^2-i\varepsilon \right)}~,
}
which describes the propagation of the shape-deformations of $p$-brane without changing the center-of-mass position. 
%
We should note that similar transition amplitude and propagator were  derived in Refs.~\cite{Ansoldi:1995hp,Ansoldi:1997cw,Ansoldi:2001km,Aurilia:2002aw} based on the first-quantized Polyakov action of $p$-brane in the mini-superspace and quenched approximations. 
In the present framework, such approximations correspond to the Born-Oppenheimer and saddle-path approximations.  

\

Finally, let us discuss the conditions under which the Born-Oppenheimer approximation would be valid.  
Equation~(\ref{approximate amplitude}) leads to the following effective action for $\mathrm{Vol}[C_p^{}(t)]$:
\aln{
S_\mathrm{eff}^{}=\int_0^{A}dt\left[k_\mathrm{Vol}^{}(t)\dot{\mathrm{Vol}}[C_p^{}(t)]-\frac{1}{2}\left(\frac{k^2}{M_p^{}(t)^2}+D[C_p^{}(t)]k_\mathrm{Vol}^{}(t)^2\right)
\right]~,\quad M_p^{}(t)=T_p^{}\mathrm{Vol}[C_p^{}(t)]~. 
}
%
One can see that the center-of-mass kinetic energy $k^2/M_p^{}(t)^2$ plays a role of a potential for the volume dynamics. 
In order to compare several time-scales, let us use the dimensionless unit such that  every quantities is normalized by an appropriate power of the $p$-brane tension $T_p^{}$. 
Then, a typical (dimensionless) time-scale of the volume-dynamics $\tau_{\mathrm{Vol}}^{}$ at a volume point $\mathrm{Vol}[C_p^{}]=V_p^{}$ can be roughly estimated as 
\aln{
\tau_{\mathrm{Vol}}^{} \sim T_p^{}\left(\bigg|D[V_p^{}]\frac{(k^2/M_p^{2})''}{k^2/M_p^{2}}\bigg|\right)^{-\frac{1}{2}}\sim T_p^{}\sqrt{\frac{V_p^2}{D[V_p^{}]}}\sim T_p^{}V_p^{\frac{p+1}{p}}~,
\label{typical frequency}
}
where $'$ denotes the derivative with respect to $V_p^{}$. 
Here, we have used a naive dimensional analysis $D[V_p^{}]\sim V_p^{-2/p}$ as it depends only on the volume-form of $C_p^{}$.~(See the definition~(\ref{def D}).) 
On the other hand, the typical diffusion time-scales of the center-of-mass and area-elements can be read from Eq.~(\ref{approximate amplitude}) as  
\aln{
\tau_{\mathrm{CM}}^{}\sim 
\tau_{\mathrm{AE}}^{}\sim AT_p^2~,
}
which implies that the volume-dynamics is negligible as long as \aln{
s\coloneq \frac{AT_p^{}}{V_p^{}}\ll L_p^{}\coloneq V_p^{\frac{1}{p}}~,
}
 where $s$ can be interpreted as the proper-time and $L_p^{}$ corresponds to the typical size of $p$-brane. 
Namely, the Born-Oppenheimer approximation seems to be invalid after the proper-time becomes comparable to the size of $p$-brane itself. 
In this sense, the point-particle limit~(\ref{point-particle limit}) appears to be always invalid, wheres the large-brane limit~(\ref{large brane limit}) remains meaningful. 
%
%
A more dedicated study is necessary to understand the propagation of $p$-brane beyond the Born-Oppenheimer approximation.

\subsection{Hausdorff dimension}

Hausdorff dimension is a measure of fractal dimension of general     objects~\cite{Gneiting_2012}. 
In the case of closed $p$-brane, it can be generally defined based on the random-surface picture as follows. 
In a discretized Euclidean space such as a cubic lattice with a lattice size $a$, let us consider a random motion of $p$-brane such that one random step is described by adding a $p$-plaquette to a given $p$-brane as illustrated by the orange block in Fig.~\ref{fig:area-derivative}.  
After $A$ steps, we end up with another $p$-brane $C_p^{}$ starting from an initial $p$-brane $C_p^{'}$. 
%
   %
Then, we count the number of triangulations constructed by $A$ $p$-plaquettes with an initial $p$-brane $C_p^{'}$ and final $p$-brane $C_p^{}$ as 
\aln{n_\chi^{}(C_p^{},C_p^{'};A)~,}
where $\chi$ represents a topology of the random surfaces.  
%
%
%
%
Then, the total number of $A$-step triangulations starting from an initial $p$-brane $C_p^{'}$ is given by 
\aln{
N_\chi^{}(C_p^{'};A)\coloneq \sum_{C_p^{}} n_\chi^{}(C_p^{},C_p^{'};A)~,
}
where the summation is taken over all $p$-branes in Euclidean spacetime.  
This allows us to define the probability distribution of finding a $p$-brane $C_p^{}$ starting from an initial $p$-brane $C_p^{'}$ as 
\aln{
P_\chi^{}(C_p^{},C_p^{'};A)\coloneq \frac{n_\chi^{}(C_p^{},C_p^{'};A)}{N_\chi^{}(C_p^{'};A)}~,
\label{probability distribution of brane}
}  
by which we can define the {\it mean-square size} of random surfaces as\footnote{
The definition of the mean-square size is not unique. 
For example, a naive one would be~\cite{Distler:1988jv,Ambjorn:1997jf,Ambjorn:2013lza}
\aln{
\sum_{C_p^{}}P_\chi^{}(C_p^{},0;A)\frac{1}{\mathrm{Vol}[C_p^{}]}\int_{S^p} d^p\xi \sqrt{h(\xi)}X^\mu(\xi)X_\mu^{}(\xi)~,
}
which requires the information of the two-point function $\langle X^\mu(\xi)X^\nu(\xi')\rangle$ on the world-sheet.  
On the other hand, we here choose the one that is constructed by the target-space quantities, $x_\mathrm{CM}^\mu$ and $\sigma^{\mu_1^{}\cdots \mu_{p+1}^{}}[C_p^{}]$. 
} 
\aln{
\langle X^2\rangle_A^{}\coloneq \sum_{C_p^{}}P_\chi^{}(C_p^{},0;A) \left(x_\mathrm{CM}^{\mu}x_{\mathrm{CM}\mu}^{}+(\sigma[C_p^{}]\cdot \sigma[C_p^{}])^{\frac{1}{p+1}}
\right)~,
}
where we have taken the vanishing limit of the initial $p$-brane for simplicity.  
When this mean-square size behaves as    
\aln{\langle X^2\rangle_A^{}\sim A^{\frac{2}{D_\mathrm{H}^{}}}\quad \leftrightarrow \quad A\sim \left(\sqrt{\langle X^2\rangle_A^{}}\right)^{D_\mathrm{H}^{}}
\label{size and volume}
}
in the large $A$ limit, $D_\mathrm{H}^{}$ represents the effective dimension of random surfaces, i.e. the Hausdorff dimension.  
%
%
Thus, we can in principle calculate $D_\mathrm{H}^{}$ once a theory/model of random surfaces is given.

In the present brane-field theory, the Euclidean transition amplitude can be identified with the provability distribution as 
\aln{
P_\chi^{}(C_p^{},C_p^{'};A)=\langle C_p^{}|e^{-\hat{H}A}|C_p^{'}\rangle~, 
}
and the result~(\ref{approximate amplitude}) in the Born-Oppenheimer approximation leads  
\aln{
\langle X^2\rangle_A^{}\propto \begin{cases}
 A & \text{(Point-particle limit)}
\\
 A^{\frac{1}{p+1}}&  \text{(Large-brane limit)}
\end{cases}
\label{central-limit theorem}
}
%
Then, from the definition of the Hausdorff dimension~(\ref{size and volume}), we obtain 
\aln{
D_\mathrm{H}^{}\approx \begin{dcases} 2  & (\text{Point-particle limit})
\\
2(p+1) & (\text{Large-brane limit})
\end{dcases}
\label{Hausdorff dimension of brane}
} 
in the Born-Oppenheimer approximation.      
In particular, the Hausdorff dimension of random $1$-brane is $D_\mathrm{H}^{}|_{p=1}^{}=4$, as conjectured by Parisi in Ref.~\cite{Parisi:1978mn}. 
%
\begin{figure}
    \centering
    \includegraphics[scale=0.7]{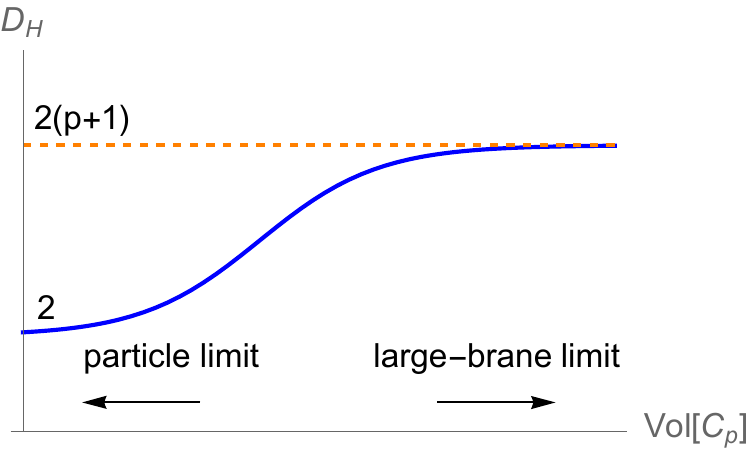}
    \caption{
  The Hausdorff dimension of $p$-brane as a function of its volume within the Born-Oppenheimer approximation.     
   }
    \label{fig:triangulation}
\end{figure}
%
Although it is not easy to examine the intermediate volume region analytically, we expect that $D_\mathrm{H}^{}$ is smoothly connected between $2$ and $2(p+1)$ as illustrated in Fig.~\ref{fig:triangulation}.  
Note that these results are valid only in the present brane-field model~(\ref{brane-field action}) and within the Born-Oppenheimer approximation.   
In fact, the Hausdorff dimension of non-critical closed string has been   calculated in many literatures~\cite{Distler:1988jv,Kawai:1991qv,Kawai:1993cj,Ambjorn:1997jf,Ambjorn:2013lza}, yielding various predictions due to different  definitions of the effective size. 
For example, in Ref.~\cite{Distler:1988jv}, the Hausdorff dimension is found to be dependent on the spacetime dimension (central charge) as $D_\mathrm{H}^{}=-2/\gamma(D)$, where $\gamma(D)$ is the string susceptibility $\gamma(D)$.    

\section{Summary}\label{sec4}
We have studied phases and propagation of closed $p$-brane in the effective field theory with higher-form global symmetries. 
Compare to the previous studies~\cite{Hidaka:2023gwh,Kawana:2024fsn,10.1093/ptep/ptaf023,Iqbal:2021rkn}, we have included the kinetic term of the center-of-mass motion as well as the kinetic term constructed by the area derivatives, which then gives rise to a new dynamical scalar (NG) mode of the $0$-form global symmetry.  
Within the mean-field analysis, we have shown that the classical brane-field exhibits the area law in the unbroken phases in the large-brane limit and that the effective theory is described by the simple Maxwell theory~(\ref{effective theory}) when the higher-form global symmetries are $\mathrm{U}(1)$. 
On the other hand, when these $\mathrm{U}(1)$ global symmetries are explicitly broken down to discrete ones, the effective theory in the broken phase is found to be a topological field theories such as Eq.~(\ref{BF theory}) and (\ref{TAE}) with multiple (emergent) higher-form discrete symmetries, resulting in topological order in general.  
%
%

%
In addition to the mean-field analysis, we have investigated the propagation of $p$-brane in the present framework. 
Compared to the naive kinetic term employing the conventional functional derivatives, our brane-field model can be more intuitively analyzed due to (i) the separation of kinetic terms between the center-of-mass and relative motions and (ii) nice geometrical properties of the area derivatives. 
We have found the plane-wave solutions of the kinetic terms and obtained the path-integral expression of the brane propagator in the Schwinger's proper time formalism.    
However, due to the mixture of the center-of-mass and relative motions, it is still challenging to perform the path-integral exactly.  
As a first step toward understanding the propagation of $p$-brane, we have applied the Born-Oppenheimer approximation by treating the brane volume as constant and obtained the approximate analytic expressions of the propagator: In the point-particle limit, the propagator reduces  to the ordinary propagator of relativistic particle with the mass $M_p^{}=T_p^{}V_p^{}$, while it describes the propagation of the area elements in the large-brane limit.   
Moreover, we have examined the Hausdorff dimension of $p$-brane and found that it varies from $2$ to $2(p+1)$ as the $p$-brane volume increases.  
While these results are intriguing, the Born-Oppenheimer approximation seems to be invalid for the point-particle limit, suggesting that the quantum nature of $p$-brane is essentially different from   relativistic particle for any parameter values.  
A more dedicated analysis is left for future investigations.

\section*{Acknowledgements}
We thank Hikaru Kawai and Yoshimasa Hidaka for the fruitful comments. 
This work is supported by KIAS Individual Grants, Grant No. 090901.   
%

\appendix

\section{Functional Derivative}\label{app:Functional Derivatives}
In this Appendix, we summarize the basics of ordinary functional derivatives. 
Let us consider a functional $F[\phi]$ of a scalar field $\phi(X)$. 
A naive functional derivative at a point $X$ can be defined by
\aln{
\frac{\delta F[\phi]}{\delta \phi(X)}=\lim_{\varepsilon\rightarrow 0}\frac{F[\phi(Y)+\varepsilon \delta^{(d)}(Y-X)]-F[\phi]}{\varepsilon}~,
\label{naive functional}
}
which is not manifestly covariant because the delta function is not. 
Instead, we can define a manifestly covariant functional derivative by
\aln{
\frac{\delta_c^{} F[\phi]}{\delta \phi(X)}=\lim_{\varepsilon\rightarrow 0}\frac{F[\phi(Y)+\varepsilon \delta_c^{(d)}(Y-X)]-F[\phi]}{\varepsilon}~,
\label{covariant functional}
}
where
\aln{ \delta_c^{(d)}(Y-X)\coloneqq\frac{1}{\sqrt{-g(X)}}\delta^{(d)}(Y-X)
}
is the covariant delta function. 
In particular, when $F[\phi]=\phi(X)$, we have
\aln{\frac{\delta_c^{}\phi(X)}{\delta \phi(Y)}=\delta^{(d)}_c(X-Y)~,
}
and one can check that Eqs.~(\ref{naive functional})(\ref{covariant functional}) are related each other as
\aln{\frac{\delta_c^{} F[\phi]}{\delta \phi(X)}=\int d^dY\frac{\delta_c^{}\phi(Y)}{\delta \phi(X)}\frac{\delta_c^{} F[\phi]}{\delta_c^{} \phi(Y)}=\frac{1}{\sqrt{-g(X)}}\frac{\delta F[\phi]}{\delta \phi(X)}~.
} 
Then, a general functional expansion is
\aln{F[\phi+\delta \phi]&=\sum_{n=0}^\infty \frac{1}{n!}\left(\prod_{i=1}^n \int d^dX_i^{}\sqrt{-g(X)}\delta \phi(X_i^{})\right)\frac{\delta^n_c F[\phi]}{\delta \phi(X_1^{})\cdots \delta \phi(X_n^{})}
\\
&=\sum_{n=0}^\infty \frac{1}{n!}\left(\prod_{i=1}^n \int d^dX_i^{}\delta \phi(X_i^{})\right)\frac{\delta^nF[\phi]}{\delta \phi(X_1^{})\cdots \delta \phi(X_n^{})}~.
\label{functional expansion}
}

\bibliographystyle{TitleAndArxiv}
\bibliography{Bibliography}

\end{document}